\RequirePackage{fix-cm}
\documentclass[smallextended]{svjour3}       % onecolumn (second format)
\smartqed  % flush right qed marks, e.g. at end of proof
\usepackage{graphicx}
\usepackage[monochrome]{xcolor}
\usepackage{algorithm}
\usepackage{algorithmic}
\usepackage{amsmath,amssymb,amsfonts}
\usepackage{multirow,longtable}

\begin{document}

\title{Multidimensional Assignment Problem for multipartite entity resolution%\thanks{Grants or other notes
%about the article that should go on the front page should be
%placed here. General acknowledgments should be placed at the end of the article.}
}
%\subtitle{Do you have a subtitle?\\ If so, write it here}

%\titlerunning{Short form of title}        % if too long for running head

\author{Alla Kammerdiner         \and
        Alexander Semenov  \and
        Eduardo L. Pasiliao
}

%\authorrunning{Short form of author list} % if too long for running head

\institute{Alla Kammerdiner \at
              University of Florida Research and Engineering Educational Facility (REEF), Shalimar, FL \\
              \email{alla.ua@gmail.com}           %  \\
%             \emph{Present address:} of F. Author  %  if needed
           \and
           Alexander Semenov \at
               Herbert Wertheim College of Engineering, University of Florida, FL, USA   \\
              \email{asemenov@ufl.edu} 
                         \and
           Eduardo L. Pasiliao \at
              Munitions Directorate, Eglin Air Force Base, Eglin, FL   \\
              \email{elpasiliao@gmail.com} 
}

\date{Received: date / Accepted: date}
% The correct dates will be entered by the editor

\maketitle
 
\begin{abstract}
Multipartite entity resolution aims at integrating records from multiple datasets into one entity.
We derive a mathematical formulation for a general class of record linkage problems in multipartite entity resolution across many datasets as a combinatorial optimization problem known as the multidimensional assignment problem. As a motivation for our approach, we illustrate the advantage of multipartite entity resolution over sequential bipartite matching. Because the optimization problem is NP-hard, we apply 
two heuristic procedures, a Greedy algorithm and very large scale neighborhood search, to solve the assignment problem and find the most likely matching of records from multiple datasets into a single entity. We evaluate and compare the performance of these algorithms and their modifications on synthetically generated data. %Moreover, we study the performance of the algorithms in a case of user profile matching with tripartite entity resolution on realistic large dataset of highly cited researchers. Importantly, our results demonstrate the capacity of the proposed optimization-based approach to multipartite entity resolution to correctly associate records from multiple datasets.
{\color{blue}We perform computational experiments to compare performance of recent heuristic, the very large-scale neighborhood search, with a Greedy algorithm, another heuristic for the MAP, as well as with two versions of genetic algorithm, a general metaheuristic.}  Importantly, we perform experiments to compare two alternative methods of re-starting the search for the former heuristic, specifically a random-sampling multi-start and a deterministic design-based multi-start. We find evidence that design-based multi-start can be more efficient as the size of databases grow large.
In addition, we show that very large scale search, especially its multi-start version, outperforms simple Greedy heuristic. Hybridization of Greedy search with very large scale neighborhood search improves the performance. Using multi-start with as few as three additional runs of very large scale search offers some improvement in the performance of the very large scale search procedure. {\color{blue}Last, we propose an approach to evaluating complexity of the very large-scale neighborhood search.}

% \PACS{PACS code1 \and PACS code2 \and more}
% \subclass{MSC code1 \and MSC code2 \and more}
\end{abstract}

\section{Introduction}
\label{sec:intro}

In this paper, we describe a computationally challenging new application of the multidimensional assignment problem (MAP). We show how this application of the MAP arises naturally when simultaneously merging multiple massive datasets. Because finding global optimum for large-size instances of the MAP  remains a significant hurdle, we apply a recent metaheuristic approach, known as the very large-scale neighborhood search (VLSNS), for solving the MAPs. Due to its novelty, the VLSNS algorithm is much less studied than other approaches for the MAP. In particular, it has not been evaluated on larger-sized problem instances, with exception of limited computational experiment in~\cite{kammerdiner2017very}. Therefore, we conduct extensive numerical experiments aimed at comparison of this solution method with state-of-the-art Greedy algorithm and with the global optimal solutions given by a standard linear solver such as Gurobi. We find that this novel metaheuristic shows good performance in our numerical comparison, and therefore, can be used in the context of merging multiple datasets known as %record linkage or 
multipartite entity resolution.  

The amount of the information on the Internet increases every year. In 2005 there were 150 exabytes digital information produced; and there were already 1200 exabytes in 2010 \cite{Helbing2011}. Nearly 70\% of this content is user generated \cite{ijwet2013}. One of the enablers of rapid growth of user-generated content are social media sites. Now there are hundreds of popular social media sites in the Internet. Many individuals maintain accounts at several sites at the same time. Social media sites allow its users to maintain a public profile describing the user. Contents of the profile vary from site to site. Different sites typically concentrate on different aspects of user's identity, and as a result information presented in user profiles is presented in different ways and typically there is not much overlapping. For instance, users of LinkedIn typically provide to the site information regarding their name, education, and employment details. Nickname, name, and such details as location may be found from Twitter, and Instagram keeps photos and details of locations visited by the user; typically the user is identified by her screenname. In order to gain complete information regarding particular user, records from multiple sites should be linked into one entity. 

Record linkage problem, a process of linking records representing the same real world entity, arises in many other applications, such as linking descriptions of the same product from multiple e-commerce sites. Another example is consolidating several databases with customers' data when companies are merged. Record linkage problem is challenging, as representation of the same entity may vary across the sources, there may be little or no overlapping in the description, and some sources may contain inaccurate or even conflicting information. 

Typical solution for pairwise entity matching is to define scoring function which returns result of pairwise comparison of records, then apply this to all pairs of potential matches, and extract pairs whose scoring is maximal. This problem is modeled in \cite{riederer_linking_2016} as maximum weighted matching on a bipartite graph. Often scoring function is restricted to be boolean, i.e. it takes values either 1 or 0 if entities match or not, respectively \cite{benjelloun_swoosh_2009}. There are methods describing record linkage provided there are erroneous values in some of the records \cite{guo_2010_vldb}. Pairwise record linkage problem seeks to identify the relations between elements of two datasets; this problem may be solved by computing similarity between each pair of items in those datasets, and finding the most similar pairs. 

However, this approach would not work for comparison between three and more datasets. This is illustrated by example below.
\begin{example}
Suppose we have three datasets: $\mathcal{A} = \{a_1,a_2,a_3\}$, $\mathcal{B} = \{b_1,b_2,b_3\}$, $\mathcal{C} = \{c_1,c_2,c_3\}$. We perform pairwise comparisons between the elements using non-transitive similarity function, and get the results as displayed in Table~\ref{example1}. The goal is to find such matching between elements of $\mathcal{A}, \mathcal{B},$ and $\mathcal{C}$ so that the sum of the scores is maximized. True matching is $(a_1,b_1,c_1)$, $(a_2,b_2,c_2)$, $(a_3,b_3,c_3)$, however, as $sim(a_1,b_1) = 0.4$, sequential matching approaches may incorrectly result in matching $(a_2,b_1,c_1)$, $(a_1,b_2,c_2)$, $(a_3,b_3,c_3)$
\begin{table}[]
\centering
\caption{Example, three-partite entity matching} 
\label{example1}
\begin{tabular}{|l|l|l|l|l|l|}
 \hline
 $\mathcal{A} - \mathcal{B}$ & sim & $\mathcal{B} - \mathcal{C}$ &sim &  $\mathcal{A} - \mathcal{C}$ &  sim  \\  \hline
$(a_1,b_1)$ & 0.4 &$(b_1,c_1)$ & 1.0 & $(a_1,c_1)$ & 1.0 \\  \hline 
$(a_1,b_2)$ & 0.6 & $(b_1,c_2)$ & 0.1 &$(a_1,c_2)$  & 0.1 \\  \hline
$(a_1,b_3)$ & 0.6 & $(b_1,c_3)$ & 0.1 &$(a_1,c_3)$   & 0.1 \\  \hline
$(a_2,b_1)$ & 0.6 &$(b_2,c_1)$  & 0.1 &$(a_2,c_1)$   & 0.1 \\  \hline
$(a_2,b_2)$ & 0.6 & $(b_2,c_2)$ & 1.0 &$(a_2,c_2)$ & 1.0 \\  \hline
$(a_2,b_3)$ & 0.6 &$(b_2,c_3)$ & 0.1 & $(a_2,c_3)$ & 0.1 \\  \hline
$(a_3,b_1)$ & 0.6 &$(b_3,c_1)$  & 0.1 &$(a_3,c_1)$ & 0.1 \\  \hline
$(a_3,b_2)$ & 0.6 &$(b_3,c_2)$  & 0.1 &$(a_3,c_2)$  & 0.1 \\  \hline
$(a_3,b_3)$ & 1.0 &$(b_3,c_3)$  & 1.0 &$(a_3,c_3)$  & 1.0 \\  \hline
\end{tabular}
\end{table}
\end{example}

In the current paper, we study multipartite records linkage problem by formulating it as the %through its formulation using
multidimensional assignment problem (MAP). Besides the advantage of multipartite entity  resolution over sequential bipartite, which was illustrated in Example~1, the formulation of multipartite entity  resolution using the MAP can offer an additional advantage in terms of matching precision. In fact, this formulation allows one to establish the most likely linkage of records %maximum-likelihood 
by solving the MAP to obtain its global optimum.  It is noteworthy that although multipartite entity  resolution has been discussed in~\cite{zhang_principled_2015}, to the best of our knowledge, it has not been formulated as the MAP. The MAP is a well-known NP-hard problem in combinatorial optimization. This problem arises in various economic, military, and health care applications, but most notably in the context of multi-sensor fusion and multi-target tracking. 

Our main contributions are as follows:
\begin{itemize}
  \item We formulate multipartite records linkage problem as multidimensional assignment problem (MAP).
  \item We perform experimental evaluation of  %We present 
  novel search algorithm for solving MAP, %and perform its experimental evaluation.
  including the exploration or multi-start phase in this search.
  \begin{itemize} 
  \item We experimentally show that our algorithm finds better solution than Greedy algorithm.
  \item {\color{blue}We compare our algorithm results to optimal solutions obtained by genetic algorithms and Gurobi solver.} %exact methods.
  %\item We experimentally show, that application of our algorithm to multipartite entity matching improves recall of the results.
  \item Importantly, we compare two distinct multi-start schemes (a random multi-start versus a design-based multi-start) for the exploration phase, and find computational evidence in support of the latter. 
  \end{itemize}
\end{itemize} 

{\color{blue}The remainder of this paper is organized as follows. 
Section~\ref{sec:lit} positions our paper in the related research literature and discusses the key contributions to both operations research and computer science.
In Section~\ref{sec:problem}, we first present intuition behind our approach to formulating  general record linkage problems for multipartite entity resolution as the MAP, and then we show how the problem statement is mathematically derived.
In Section~\ref{sec:sol-algs}, we outline two algorithms developed specifically for solving the MAP and focus our attention on a newer, VLSN-based approach. For completeness, we include pseudocode of each algorithm. First, we briefly describe Greedy algorithm. Second, we present the VLSN local search algorithm. Third, we propose several multi-start strategies for the VLSN search and discuss the resulting VLSN-based metaheuristics. 
Section~\ref{sec:complexity} presents our approach to evaluating complexity of the VLSN-based metaheuristics using two related Markov chains.
In Section~\ref{sec:exp}, we study the performance of VLSN-based metaheuristics in terms of the solution quality and compare VLSN with non-VLSN algorithms, including Greedy algorithms, standard metaheuristic such as genetic algorithm, and an all-purpose solver such as Gurobi. Additionally, we illustrate the application of our approach to real record linkage and multipartite resolution on synthetic dataset.
Finally, Section~\ref{sec:end} concludes.
}

\section{Related research}\label{sec:lit}
The paper builds upon several strands of the literature in mathematical programming and computer science.  On one side, the paper is concerned with the multidimensional assignment problem in combinatorial optimization. On the other side, the study advances record linkage, entity resolution, and other related areas in relational databases. In this section, we review the prior research in these areas, starting first with computer science and then continuing with optimization.

In computer science, related areas include research on record linkage, network alignment, user identity reconciliation, de-anonymization of online social networks, entity resolution, and reference reconciliation. 
Survey \cite{brizan_survey_2006} reports entity resolution and record linkage methods used in relational databases. The survey focuses on criteria for entity matching (such as exact match, distance match, etc); survey considers pair-wise linkage only. Algorithm for linking user accounts across two social media sites is presented in the article \cite{riederer_linking_2016}. Authors model results produced by users at different sites as trajectories (these trajectories are modeled as buckets of records), compute the scores for every pair of users, represent this data as bipartite graph and find the maximum weighted matching on this graph. Authors consider linking across two sites. Review of user identity linkage across social networks \cite{kai_shu_2016} concentrates on linkage of user profiles between two social networks (source site and target site). Linking users between two social media sites by comparing usernames is presented in \cite{icwsm_2009}. Authors of \cite{he_relationship_2016} construct relationship between two social networks by applying decision tree classification to set of users' features (same highschool, etc).

According to \cite{Papadakis_VLDB_2016}, entity resolution (identifying entities, that
represent the same real-world object) has quadratic complexity, and because of that many algorithms employ blocking: cluster similar elements into blocks, and then perform cluster-wise comparison. Paper presents comprehensive survey of blocking methods; it focuses on resolution of entities over two datasets. Evaluation of entity resolution approaches is presented in \cite{Kopcke_VLDB_2010}. Authors consider real world datasets, and several algorithms for evaluation, however matching is carried out between two datasets only. Linking temporal entities is described in \cite{Li_sigmod_2015}. Paper concentrates on building temporal model of user profile evolution, and then provides matching algorithm across two datasets. Paper \cite{sagi_sigmod_2016} describes how fuzzy clustering may be used as a blocking method; authors applied the entity resolution methods to the repository of
Holocaust-era information. 
%%%%%%%%%%%%%%%%%%%%%%%%%%%%%%%%%%%%%%%%%%%
%%%%%%%%%%%%%%%%%%%%%%%%%%%%%%%%%%%%%%%%%%%

Paper \cite{zhang_principled_2015} describes entity resolution problem as combinatorial optimization for problem of max-weight graph matching on multi-partite graphs. Paper presents optimization model and several algorithms to solve it. Multipartite entity resolution problem is also studied in \cite{Ye_CIKM_2015}. Online version of entity resolution with an oracle which correctly labels matching and non-matching pairs through queries is studied in \cite{firmani_online_2016}. Distributed algorithms for data deduplication are studied in \cite{Chu_VLDB_2016}. \cite{puglisi_web_2016} describes how trackers track user in the Internet and further apply the information for personalized advertising. Papers \cite{Gokhale_sigmod_2014},\cite{wang_crowder_2012} describe application of crowdsourcing to entity matching. Paper focuses on application of blocking, and minimization of number of comparisons. Identification of Facebook users in Twitter is presented in \cite{Jain_2013}. Paper \cite{Zhou_2016} describes how data on adjacent users in the social network may be used for matching of user profiles between two social networks. Alignment of multiple social networks is studied in \cite{zhang_multiple_2015}. Survey of automated algorithms for duplicate detection and various similarity measures is presented in \cite{elmagarmid_duplicate_2007}. Authors of \cite{benjelloun_swoosh_2009} view record comparison functions and black-boxes, and analyze generic approach for entity resolution. Authors propose monotonicity conditions of merge, and describe entity  resolution approach with idempotence, commutativity, associativity and representativity properties.
%%%%%%%%%%%%%%%%%%%%%%%%%%%%%%%%%%%%%%%%%%%%

In combinatorial optimization, the multidimensional assignment problem (MAP), sometimes called the multi-index assignment, %is an NP-hard problem that
was first introduced in~\cite{pierskalla1967tri,pierskalla1968letter}.  This problem belongs to %is a part of a larger 
a general %broader 
class of  assignment problems~\cite{burkard1999linear,burkard2012assignment}, such as the linear assignment problem (LAP) and the quadratic assignment problem (QAP). 
The MAP is shown to be NP-complete~\cite{karp1972reducibility,garey1979computers}. In its simplest form, the MAP is usually characterized by two parameters, namely cardinality $N$ and dimensionality $M$. Then the number of feasible solutions of the MAP is $(N!)^{M-1}$, growing super exponentially. Three-dimensional version of the MAP is known as the axial 3-dimensional assignment or axial 3-AP. The axial 3-AP has an alternative closely related problem, known as the planar 3-AP. 

The MAP found applications in diverse areas ranging from defense to physics. For example, the axial 3-AP can arise in the context of capital investment, dynamic facility location, satellite launching~\cite{pierskalla1967tri,pierskalla1968letter}. Furthermore, the 3-AP can arise in production planning in automated manufacturing, such as in assembly of printed circuit boards~\cite{crama1997assembly,crama2012production}. An interesting special case of the axial 3-AP with binary coefficients (i.e., zeros and ones) arises in application to perishable production planning~\cite{arbib1999three}. Applications of the related planar 3-AP include school timetabling~\cite{hilton1980reconstruction,frieze1981algorithm}, experimental design~\cite{hilton1980reconstruction}, rostering~\cite{gilbert1987algorithm,gilbert1988multidimensional}, and  launch of communication satellites~\cite{balas1983traffic}. 

The applications discussed so far lead to three-dimension versions of the MAP. A general (axial) MAP with higher dimensionality (i.e., $M>3$) can arise in relation to defense and military surveillance application in the field of multi-sensor multiple target tracking~\cite{poore1993data}. In this setting, the task is to solve a data association problem of identifying which sensor measurements or scans belong to which target paths or tracks. The axial MAP with dimensionality of $M=5$ arises in tracking charged elementary particles in colliding beam experiments in high energy physics~\cite{pusztaszeri1996tracking,pusztaszeri2000nonlinear}.  Routing in meshes or two-dimensional array of nodes (i.e., switches at a circuit level or processing elements at network interconnection level) can be formalized as an axial 3-AP with additional side constraints, which can be further reformulated into a standard sparse MAP with dimensionality $M=5$. Recently, biomedical and healthcare application of general MAP to detection and situational assessment of falls and near falls in elderly population using wearable sensors is described and studied in~\cite{kammerdiner2015ranking,kammerdiner2019data}. 

% Pierskalla, 1967; Pierskalla, 1968 mention a number of settings in which axial 3IAPs arise: capital investment, dynamic facility location, satellite launching. Further, in Crama et al., 1996 asituation in the assembly of printed circuit boards is described that is modeled using formulation (A3IAPMIN). Arbib et al., 1999 describe a problem in perishable production planning formulated as an axial 3IAP with f0; 1g coefficients.
In this paper, we contribute to this literature by introducing a new application to record linkage and entity resolution problems in multiple relational databases. The feature of the MAPs arising in this context is that a problem instance has massive cardinality $N$ and small-to-moderate dimensionality $M \in [3,10]$. Because the MAP is generally NP-hard for $M\geq 3$, a natural question is how %%%%% (finished 5/13) 
to solve the MAP instances to perform record linkage of real-life databases. Although we use a standard linear solver to get the global optimal solution, for sufficiently large $N$ even the best solver (e.g., Gurobi) will fail to return the global solution in a reasonable time. Therefore, we examine two approximate algorithm, namely (i) a promising new metaheuristic approach (VLSN search) that exploits a way to quickly search through a massive neighborhood of $N!$ solutions, and (ii) a Greedy heuristic that has good theoretical performance guarantees for asymptotic case ($N\to\infty$). Moreover, we hybridize these two heuristics by using a Greedy solution as a starting solution in the VLSN search. Furthermore, we contribute to the literature on search-based metaheuristics by exploring two alternative approaches to sampling or constructing solutions for re-initializing the search in each exploitation stages. These two alternative ways of exploration are random sampling and (deterministic) ``design''-based constructive heuristic. %While design-based methods are employed in 
{\color{blue}Additionally, we contribute to the literature on metaheuristics for the MAP by comparing the VLSN-based algorithms with two versions of genetic algorithm (GA). Our results demonstrate that VLSN search tends to outperform GA on moderate-to-large instances.
Last, %Last but not least,
because little is known about complexity of VLSN search, we propose an approach for evaluating worst-case complexity of the VLSN metaheuristics. }
%%%%%%%%%%%%%%%%%%%%%%%%%%%%%%%%%%%%%%%%%%%%
\section{Assignment Formulation of General Record Linkage Problems}\label{sec:problem}
In this section, we formulate a general class of record linkage problems in the context of multipartite entity resolution as the multidimensional assignment problem (MAP), building upon the work of~\cite{zhang_principled_2015}. We discuss variations on this formulation in the conclusion of this section.

%%%%%%%%%%%%%%%%%%%%%%%%%%% START COLOR BLUE 
{\color{blue} %edede
To clarify the problem statement, we first present general intuition behind formulation of multipartite record linkage as the MAP. %assignment 
Recall that multipartite record linkage problem arises when merging records in multiple datasets, where each of these datasets contains different aspects of underlying entities. In application to user profile matching, linkage problem is matching records from multiple social media sites (e.g., Google, LinkedIn, Twitter) to associate these records with a user (i.e., entity). To formulate the problem, we assume that media site records have been deduplicated so that each record is unique. Next, we define a similarity function on the $M$-dimensional vector of records where each dimension corresponds to a given social media site (i.e., a given dataset included in a merge). The task is to partition the merged dataset into the $M$-dimensional vectors of records. For example, suppose the record linkage is performed with three records of Google accounts $g_1,g_2,g_3$, two LinkedIn accounts (or records) $l_1,l_2$, and two Twitter accounts (or records) $t_1,t_2$. Using similarities among accounts, the most likely matching returns three user profile vectors, namely user profile~1 with $(g_1,l_2,\cdot)$, profile~2 with $(g_2,\cdot,t_2)$ and profile~3 with $(g_3,l_1,t_1)$, where $\cdot$ represents an empty record or an absence of an account on a given site.
The key intuition behind our approach is that the
records from $M$ sites are partitioned into the $M$-dimensional vectors or tuples so as to maximize the likelihood of similar records across sites being assigned together into a tuple. Observe that in partition, at most one record from each site is assigned to a given tuple.
}%%%%%%%%%%%%%%%%%%%%%%% END COLOR BLUE

Let $E \in \mathcal{E}$ be a real-world entity belonging to set of entities $\mathcal{E}$. %Each entity is described by a set of fields $\mathcal{F}_E$. %% $\mathcal{F}$.
Entities are represented by records in a source or a dataset. \textcolor{blue}{Entities are not observed directly. They are represented by their correct association of records from multiple sources.} We assume that if an entity exists in a source then the entity is uniquely represented by a record. In other words, records in each source or dataset are deduplicated. Let a common set $\mathcal{R}$ of all possible records $R$ consist of representations of an entity $E \in \mathcal{E}$ in some source or dataset $D$. Alternatively, for any record $R$ found in some source $D$, we have $R \in \mathcal{R}$.

Let $M$ be the total number of sources or datasets that must be integrated. \textcolor{blue}{This integration is done by matching the records of each source or dataset with the corresponding records that represent the same unknown entity in other sources.} $M$ is an integer such that $M\geq 2$. For each $k=1,2,\ldots,M$, let $D(k)$ denote a set that consists of a dummy or missing records and $N_k$ actual records (which are unique or deduplicated in the $k$-th dataset). Suppose that non-negative integers $i_k=0,1,\ldots,N_k$  denote the indexes of records in the $D(k)$-th dataset (or source) with the index $i_k=0$ representing the missing (or dummy) record, which has no field entries. 

Let $sim (\cdot,\cdot): \mathcal{R} \times  \mathcal{R} \mapsto [0,1]$ be a similarity function that compares any pair of records $R_1,R_2 \in \mathcal{R}$, mapping them to their similarity with dissimilar records represented by 0 and similar ones denoted by 1. %Let $sim (\mathcal{E} \times  \mathcal{E} ) \mapsto [0,1]$ be a similarity function, mapping pair of entities to their similarity, 0 meaning dissimilar, and 1 meaning similar.
\begin{definition}
 Given datasets $D(1),D(2), \ldots, D(M)$, $M$ corresponding records $R_1,\ldots,R_M$, and a pair-wise similarity function $sim (\cdot,\cdot): \mathcal{R} \times  \mathcal{R} \mapsto [0,1]$, we define a \emph{multi-record} similarity function $msim(\cdot,\ldots,\cdot): \mathcal{R}^M \mapsto [0,1]$ as follows:
\begin{equation}
msim(R_1,\ldots,R_M) = \sum_{i,j:i<j} sim(R_i,R_j).
\end{equation}
\end{definition}
In addition, let us consider a related function $\delta$: 
\begin{eqnarray} \label{eq:delta}
\delta(R_1,\ldots,R_M) &:=& \sum_{i,j:i<j} \left(1-sim(R_i,R_j)\right) \\
\nonumber
&=& (M-1)M/2 - msim(R_1,\ldots,R_M). 
\end{eqnarray}

%Let $F_k$ denote the number of fields $F^{k} \in \mathcal{F}$ in the dataset (source) $k$. We note that the field indexes $f^k=1,2,\ldots,F_k$ can be re-odered in such a way as to ensure a correspondence of $F$ common fields $f=1,\ldots,F$ either for pairwise comparison ($f^k=f^j=f$ for some $k\neq j$) or multisource comparision ($f^{k_1}=f^{k_2}=\ldots=f^{k_r}=f$ for some $k_1\neq k_2 \neq \ldots k_r$) of fields. So, we assume schema matching problem to be solved. In general, the number of common fields, $F$, can vary depending on a specified comparison \cite{brizan_survey_2006} \cite{benjelloun_swoosh_2009}.%%%For a given pairwise comparison of different databases $k$ and $j$, let the number of common fields between two databases be $F(k,j)=F(j,k)=F$, and let nonpositive integers $i_k=a$ and $i'_j=b$ such that $a<F$ and $b<F$, respectively, denote two compared entities: the entity $a$ from database $k$ and the entity $b$ from database $j$. We denote $P(a\leftrightarrow b)$

Furthermore, let $d^{k}_{i_k}$ represent the $i_k$-th record in the $k$-th dataset for some $k=1,2,\ldots,M$ and $i_k=0,1,\ldots,N_k$. Then for any given $k$, $d^{k}_{0}$ is the missing (i.e., dummy) record. Also 
\begin{equation}
 D(k) = \cup_{i_k=0}^{N_k}\{d^{k}_{i_k}\}, \textrm{ for any }k=1,\ldots,M.
\end{equation}
We combine $M$ sets $D(1),D(2),\ldots,D(M)$ of entities into the cumulative collection of $M$ datasets, written as
\begin{equation}
D^{M}:=\{D(1),\ldots,D(M)\} = \cup_{k=1}^{M}D(k).
\end{equation}
\textbf{Problem Definition.} In a general %$M$-
multipartite entity resolution, the goal is to link records from multiple sources (datasets) that corresponds to the same {\color{blue}unknown} entity. Hence, the decision making problem is to find the most likely {\it partition} of all possible $M$-tuples from the cumulative dataset $D^M$ into linkages (or $M$-partite matches) and non-linkages, given a collection of $M$ databases.

We define a linkage (or $M$-partite matching) of records $D_{i_1 \ldots i_M}$ as
\begin{equation}\label{eq:DecVar}
D_{i_1 \ldots i_M} := \{d^1_{i_1},\ldots,d^M_{i_M}\}, 
\end{equation}
where exactly one record $d^k_{i_k}$ is included from the dataset $D(k)$ for each $k=1,\ldots,M$. According to this definition, we have $\color{blue}D_{0\ldots 0}:= \{ d^1_0,\ldots,d^M_0 \}$. %\{ z^1_0,\ldots,z^M_0 \}$. 

Next, we define a set $\Gamma$ of all feasible partitions $\gamma$ of $D^{M}$ that satisfy the following conditions (or constraints):
\begin{eqnarray}\label{eq:linkage}
\gamma &=& \{\gamma_1,\ldots,\gamma_{m(\gamma)}\} \\ 
\label{eq:empty}
\gamma_i \cap\gamma_j &\subset& D_{0\ldots 0}, \textrm{ for }i\neq j  \\ 
\label{eq:AllSources}
D^{M} &=& \{d^1_0,\ldots,d^M_0\} \cup \left( \cup_{j=1}^{m(\gamma)} \gamma_j\right) \\
\label{eq:MissingNotLinked}
\gamma_i &\neq & \{d^1_0,\ldots,d^M_0\} \; \forall j=1,\ldots,m(\gamma)\\
\label{eq:corresp}
\forall \gamma_j: \gamma_j &=& D_{i_1\ldots i_M} \; \exists !\; M\textrm{-tuple }(i_1,\ldots,i_M)
\end{eqnarray}

{\color{blue} These conditions have the following meanings. Eq.~\ref{eq:linkage} states that each feasible partition $\gamma$ consists of $m(\gamma)$ record linkages $\gamma_i$ that match records among $M$ datasets.  
 Eq.~\ref{eq:empty} implies that any two distinct linkages only share missing records.
 Eq.~\ref{eq:AllSources} says that the cumulative collection $D^M$ of $M$ datasets or sources  is comprised of record linkages forming a partition and missing records from each source.
 Eq.~\ref{eq:MissingNotLinked} states that no linkage matches only missing records. 
 Importantly, Eq.~\ref{eq:corresp} guarantees an existence of a unique ordered set of the positions $i_m$ of records from each source $m = 1,\ldots, M$ for each linkage or matching $\gamma_j, j=1,\ldots,m(\gamma)$ of records from a partition $\gamma$. Intuitively, together the conditions insure that a partition covers all the combined data from $M$ sources by splitting these data into the $M$-tuples $(i_1,\ldots,i_M)$ of records, which represent a unique hypothesized entity, and that $M$-tuples are distinct up to missing record(s) for some sources.
}

A missing record $d^{k}_{i_k}$ is included in the above definition, and we denote it by $D_{0\ldots 0 i_k 0\ldots 0}$. %Recall that $N_k$ was defined as the number of actual entities or deduplicated records in the $k$-th database (source). 
Using~\eqref{eq:DecVar} --~\eqref{eq:corresp}, we replace $\gamma_i$'s with the respective linkage variables $D_{i_1\ldots i_M}$ to rewrite the conditions satisfied by the set $\Gamma$ of all feasible partitions $\gamma$ of $D^{M}$ as follows:
\begin{eqnarray}\label{eq:linkage.D}
\gamma &=& \left\{D_{i_1\ldots i_M} \textrm{meets~\eqref{eq:empty.D} -- \eqref{eq:MissingNotLinked.D},} \right. \\ 
\nonumber
&& \left. \textrm{for }i_k=0,\ldots,N_k, k = 1,\ldots,M\right\} \\
\label{eq:empty.D}
D_{i_1\ldots i_M} \cap D_{j_1\ldots j_M} &\subset& D_{0\ldots 0}, \\
\nonumber & &\textrm{ when }\exists k\in\{1,\ldots,M\}: i_k\neq j_k  \\ 
\label{eq:AllSources.D}
D^{M} &=& D_{0\ldots 0} \cup \left( \bigcup_{D_{i_1\ldots i_M}\in\gamma} D_{i_1\ldots i_M}\right) \\
\label{eq:MissingNotLinked.D}
D_{0\ldots 0} &\notin & \gamma %\\ 
%\label{eq:corresp.D}
%\forall \gamma_j: \gamma_j &=& D_{i_1\ldots i_M} \; \exists !\; M\textrm{-tuple }(i_1,\ldots,i_M)
\end{eqnarray} 

The above conditions~\eqref{eq:empty.D} and~\eqref{eq:AllSources.D} are equivalent to the condition that each actual record $d^{k}_{i_k}$ is included in only one record linkage $D_{i_1\ldots i_M} \in \gamma$. We formalize this condition by introducing the 0-1 decision variables:
\begin{equation}\label{eq:vars}
 x_{i_1\ldots i_M} = 
 					\left\{ 
                    	\begin{array}{cc}
                    		1, & \textrm{if }D_{i_1\ldots i_M}\in\gamma;\\
                            0, & \textrm{otherwise.}
                    	\end{array}
 					\right.
\end{equation}
This leads to the following equivalent characterization of the feasible partitions in $\Gamma$ in terms of the new variables:
\begin{eqnarray}\label{eq:constraint.MAP}
	\sum_{(i_1,\ldots, i_{k-1},i_{k+1},\ldots, i_M)=(0,\ldots,0)}^{(N_1,\ldots, N_{k-1},N_{k+1},\ldots, N_M)} x_{i_1\ldots i_M} = 1, \\
    \nonumber\textrm{ for }i_k=1,\ldots,N_k,\; k=1,\ldots,M.
\end{eqnarray}

The decision problem of record linkage in the context of multipartite entity resolution is concerned with choosing a feasible partition $\gamma\in\Gamma$ that maximizes the posterior probability $P(\gamma|D^M)$ given the data, i.e.,
\begin{equation}\label{eq:Max}
	\max_{\gamma\in\Gamma} P(\gamma|D^M).
\end{equation}
Note that because multiplying the objective by the constant (e.g., $\left[P(\gamma^0|D^M)\right]^{-1}\equiv \textrm{const}$ for specific $\gamma^0$) does not change the solution of the optimization problem, \eqref{eq:Max} is equivalent to 
\begin{equation}\label{eq:MaxMissing}
	\max_{\gamma\in\Gamma} \frac{P(\gamma|D^M)}{P(\gamma^0|D^M)},
\end{equation}
where $\gamma_0$ denote a partition that consists of all missing records. In terms of $D_{i_1\ldots i_M}$, the partition $\gamma^0$ can be written as
\begin{equation}\label{eq:gamma0}
	\gamma^0 = \left\{ D_{0\ldots 0 i_k 0\ldots 0} |i_k=1,\ldots,N_k,\; k=1,\ldots,M \right\}.
\end{equation}
Using Bayes' theorem:
\begin{equation}\label{eq:Bayes}
	 P(A|B)=\frac{P(B|A)P(A)}{P(B)},  \forall\textrm{ events }A,B : P(B)\neq 0,
\end{equation}
we can write the posterior in~\eqref{eq:Max} and~\eqref{eq:MaxMissing} as:
\begin{equation}\label{eq:posterior}
 P(\gamma|D^M) = \frac{g(D^M|\gamma)P(\gamma)}{g(D^M)},
\end{equation}
where $g(D^M)$  and $g(D^M|\gamma)$ denote the joint probability density function of the observed data $D^M$ in $M$ databases and the conditional probability density function of the observed data $D^M$ given a partition $\gamma$, respectively. Note that partitions have a discrete distribution, while the observed data are supposed (for simplicity) to be continuous (although the latter could be written as mixture of discrete and continuous variables if needed).

It is reasonable to assume that given a partition individual record linkages are mutually independent. Hence, the likelihood can be written as:
\begin{equation}\label{eq:likelihood}
 g(D^M|\gamma) = \prod_{D_{i_1\ldots i_M}\in\gamma} g(D_{i_1\ldots i_M}|\gamma),
\end{equation}
whereas the prior can be written as:
\begin{eqnarray}\label{eq:prior}
 P(\gamma) &=& \left(\int g(D^M) dD^M \right)\times \\ 
 \nonumber 
 &\times &\prod_{D_{i_1\ldots i_M}\in\gamma} \int g(D_{i1\ldots i_M}|\gamma) dD_{i1\ldots i_M}.
\end{eqnarray}
Using~\eqref{eq:likelihood} and~\eqref{eq:prior}, the posterior~\eqref{eq:posterior}
can be rewritten as
\begin{eqnarray}\label{eq:posteriorLong}
&& P(\gamma|D^M) = \frac{\int g(D^M) dD^M }{g(D^M)}\times \\ 
 \nonumber 
&&\times \prod_{D_{i_1\ldots i_M}\in\gamma} g(D_{i_1\ldots i_M}|\gamma)\cdot\left(\int g(D_{i1\ldots i_M}|\gamma) dD_{i1\ldots i_M}\right)
\end{eqnarray}
Additionally, if $\gamma,\rho\in\Gamma$ are two distinct partitions (i.e., $\gamma\neq\rho$) and $D_{i_1 \ldots i_M} \in \gamma \cap \rho$, we assume that the following holds:
\begin{eqnarray}\label{eq:UniqueProd}
	g(D_{i_1\ldots i_M}|\gamma)\cdot\left(\int g(D_{i1\ldots i_M}|\gamma) dD_{i1\ldots i_M}\right) = \\
    \nonumber 
    = g(D_{i_1\ldots i_M}|\rho)\cdot\left(\int g(D_{i1\ldots i_M}|\rho) dD_{i1\ldots i_M}\right).
\end{eqnarray}
We can write the objective function in~\eqref{eq:MaxMissing} as follows, using~\eqref{eq:posteriorLong}:
\begin{eqnarray}\label{eq:Obj}
	&&\frac{P(\gamma|D^M)}{P(\gamma^0|D^M)} =  \\
    \nonumber
    &&= \prod_{D_{i_1\ldots i_M}\in\gamma} \frac{(D_{i_1\ldots i_M}|\gamma)}{J_0}\times\left(\int g(D_{i1\ldots i_M}|\gamma) dD_{i1\ldots i_M}\right),
\end{eqnarray}
where
\begin{equation}
	J_0 = \prod_{k=1,
    i_k\neq 0}^{M} g(D_{0\ldots i_k \ldots 0}|\gamma^0)\left(\int g(D_{0\ldots i_k\ldots 0}|\gamma^0) dD_{0\ldots i_k \ldots 0}\right)
\end{equation}

Applying a negative natural logarithm to both sides~\eqref{eq:Obj}, we have:
\begin{eqnarray}\label{eq:MinObj}
	&& -\ln \left(  \frac{P(\gamma|D^M)}{P(\gamma^0|D^M)} \right) =  \\
    %&&\ln P(\gamma^0|D^M) -\ln P(\gamma|D^M) =  \\
    \nonumber
    &&= \sum_{D_{i_1\ldots i_M}\in\gamma} \left\{ \ln J_0 - \ln g(D_{i_1\ldots i_M}|\gamma)
    \right.\\
    \nonumber
    && 
    \left.
    -\ln \left(\int g(D_{i1\ldots i_M}|\gamma) dD_{i1\ldots i_M}\right) \right\}.
\end{eqnarray}
Setting
\begin{eqnarray}\label{eq:coef}
	c_{i_1\ldots i_M} &=& \ln J_0 - \ln g(D_{i_1\ldots i_M}|\gamma) \\
    \nonumber
     &-&\ln \int g(D_{i1\ldots i_M}|\gamma) dD_{i1\ldots i_M},
\end{eqnarray} we obtain from~\eqref{eq:MinObj}:
\begin{equation}\label{eq:LogAsSum}
	-\ln \left(  \frac{P(\gamma|D^M)}{P(\gamma^0|D^M)} \right) = \sum_{D_{i_1\ldots i_M}\in\gamma} c_{i_1\ldots i_M},
\end{equation}
and the maximum likelihood~\eqref{eq:MaxMissing} with the constraints~\eqref{eq:constraint.MAP} on partitions $\gamma\in\Gamma$ is restated as the following minimization problem, known as the multidimensional assignment or the MAP:
\begin{eqnarray}~\label{eq:MAP}
	&&\textrm{minimize } y:=\sum_{i_1=0}^{N_1}\cdots\sum_{i_M=0}^{N_M} c_{i_1\ldots i_M}x_{i_1 \ldots i_M}\\
    \nonumber
    &&\textrm{subject to} \\
    \nonumber%%%
    &&\sum_{(i_1,\ldots, i_{k-1},i_{k+1},\ldots, i_M)=(0,\ldots,0)}^{(N_1,\ldots, N_{k-1},N_{k+1},\ldots, N_M)} x_{i_1\ldots i_M} = 1, \\
    \nonumber
    &&\textrm{ for }i_k=0,1,\ldots,N_k,\; k=1,\ldots,M.
\end{eqnarray}

We %Without loss of generality, we 
assume that the cost coefficients in~\eqref{eq:coef} in the optimization problem~\eqref{eq:MAP} are proportional to the corresponding $\delta$ in~\eqref{eq:delta}. This assumption is natural for the setting of user profile matching. In fact, any cost coefficient $c_{i_1\ldots i_M}$ in~\eqref{eq:coef} corresponds to negative log-likelihood of the matching of records $(i_1,\ldots, i_M)$, where the records are ordered in accordance with datasets. For $c_{i_1\ldots i_M}$, we have $\delta(i_1,\ldots,i_M) = \sum_{k=1}^{M}\sum_{j<k} \left(1-sim(i_j,i_k)\right) = (M-1)M/2 - msim(i_1,\ldots,i_M)$, and so $\exp\{-\delta(i_1,\ldots, i_M)\} \propto msim(i_1,\ldots,i_M)$. Hence, $\exp\{-c_{i_1\ldots i_M}\} \propto msim(i_1,\ldots,i_M)$. In other words, the corresponding likelihood of matched records $(i_1,\ldots, i_M)$ is proportional to the multi-record similarity $msim(i_1,\ldots,i_M)$. Of course, the minimization of negative log-likelihood is simply a maximum likelihood approach. Naturally, in user profile matching, more similar records are more likely to represent a better match.

\section{Solution of the multidimensional assignment problem (MAP)}\label{sec:sol-algs}
The solution approaches for the MAP include a number of exact and approximate methods ranging from branch-and-bound~\cite{pierskalla1968letter} %{Pier68} 
and Lagrangian relaxation~\cite{frieze1981algorithm} %{Frie81} 
to tri-substitution~\cite{pierskalla1967tri} %{Pier67}
and such metaheuristics as tabu search, simulated annealing, greedy randomized adaptive search procedure, memetic approach, and hybridization of several techniques~\cite{kammerdiner2008multidimensional}. %{Kamm09ency}.
%Many metaheuristics utilize a \textbf{local search} of the solution space~\cite{Guti09,Kara11}.
The problem %have recently
received an increased
attention in the last decade~\cite{%grundel2005average,grundel2007number,
gutin2008worst,kammerdiner2007characteristics,kammerdiner2009application,karapetyan2011local,krokhmal2011optimality,nguyen2014solving,pasiliao2010local,%poore2006some,
vogiatzis2014graph}. Here we will utilize and discuss two solution algorithms, namely a Greedy algorithm~\cite{krokhmal2011optimality} and very large-scale neighborhood search~\cite{kammerdiner2017very}. 

\subsection{Greedy algorithm for the MAP with discrete costs}
Here we describe a greedy solution algorithm proposed in~\cite{krokhmal2011optimality} for solving the MAP with discrete costs. This algorithm works by iteratively finding the smallest cost coefficient $c_{i_1\ldots i_M}$ and then removing cost coefficients $c_{j_1 \ldots j_M}$ with $i_k = j_k$ for any $k=1,\ldots, M$ (so that infeasible solutions cannot be selected). With an exception of the notation, here we will follow Algorithm~1 in~\cite{krokhmal2011optimality}, including %with an exception of the notation and 
the assumption of equal cardinalities  $N_1=\ldots = N_M =n$ for all dimensions. %Here we allow for different cardinalities and suppose that $N_1\geq \ldots \geq N_M$ (without loss of generality) by indexing the datasets in a decreasing order of their records. 
Further research is needed to adapt this algorithm to work for a general case with different cardinalities, possibly by introducing dummy records in some dimensions to associate with partial profile matches.

\begin{algorithm}[h!]
\caption{A greedy algorithm for the MAP with discrete costs}%{Basic Local Search}
\label{alg:greedy}
\begin{algorithmic}[1]
\STATE Input $n$, $M$, %Initialize 
$C = \{ c_{i_1 \ldots i_M} | i_k=1,\ldots,n, %N_k%, 
k=1,\ldots,M \}$
\STATE Initialize $y \leftarrow 0$
\STATE Initialize $n^M$-array $X = \{ x\left(i_1,\ldots,x_M \right)\}\leftarrow 0$
\FORALL{$m \in \{1,\ldots,M\}$}
	\STATE Define set $\mathcal{N}_m \leftarrow \{1,\ldots,n%N_k
    \}$
\ENDFOR
\FORALL{$k \in \{1,\ldots,n\}$}
	\STATE Define submatrix $C^{(k)}\in \mathbb{R}^{(n-k+1)^M}$ of the matrix $C$ as
    $C^{(k)} \leftarrow \{ c_{i_1 \ldots i_M} : i_k\in \mathcal{N}_k \forall k\}$
    \STATE Find the smallest element $c_{i_1^{(k)} \ldots i_M^{(k)} }$ of $C^{(k)}$ such that
    $(i_1^{(k)} \ldots i_M^{(k)})\in \arg\min\{ c_{i_1 \ldots i_M } \in C^{(k)} \}$
    \STATE Let $y \leftarrow y+ c_{i_1^{(k)} \ldots i_M^{(k)} }$
    %moved line 9 in Krokhmal's Greedy  here
    \STATE Let $x\left(i_1^{(k)}, \ldots, i_M^{(k)} \right) \leftarrow 1$
    \FORALL{$m \in \{1,\ldots,M\}$}
		\STATE Update set $\mathcal{N}_m \leftarrow \mathcal{N}_m \setminus \{i_m^{(k)} \}$
	\ENDFOR
\ENDFOR
%\FORALL{$k\in\{1,\ldots,n \}$}
%	\IF{$x$}  \STATE Define $X$ \STATE $x_{i_1}$ \ENDIF
%\ENDFOR
\STATE Output a feasible solution $X$ of the MAP~\eqref{eq:MAP} and cost $y$
\end{algorithmic}
\end{algorithm}

\subsection{Very large-scale neighborhood (VLSN) local search}

We describe the solution algorithm %$$$% 
that was recently proposed in~\cite{kammerdiner2017very}. 
Specifically, we present the version of
the VLSN search algorithm for the MAP that is based on finding an optimal permutation for each dimension. %$$$% The procedure is shown in Algorithm~\ref{alg:VLSN.OP} and it corresponds to Algorithm~2 in~\cite{kammerdiner2017very}. 
The search alternates between two phases: (i) the exploitation phase (or a descend to a local minimum) and (ii) the exploration phase (or a re-initialization of a descent from a different solution).

We will follow the notation introduced in~\cite{kammerdiner2017very}. For convenience, we assume $N_1=\ldots = N_M = n$, i.e., cardinalities are the same in each dimension. The algorithm can be modified to allow for different cardinalities. For example, one can always index the datasets in a decreasing order of contained records and introduce enough dummy records into datasets with smaller cardinalities.  %$!$!$% However, that paper assumes that  $N_1=\ldots = N_M$, i.e., cardinalities are the same in each dimension. To allow for different cardinalities, we assume $N_1\geq \ldots \geq N_M$ without loss of generality. In fact, one can always index the datasets in a decreasing order of contained records.

We take advantage of a well-known alternative formulation of the MAP~\eqref{eq:MAP} using the combinatorial optimization representation of the MAP solutions as the ordered set of $M-1$ permutations of size $n$, where $M,n$ are dimensionality and cardinality of the problem (see, e.g.,~\cite{burkard1999linear}). Let $S$ denote %the current solution, i.e., 
any given feasible solution of the MAP. Solution $S$ %It 
could be some %a 
starting solution %from multi-start strategy $\Sigma$ 
or the current %a feasible 
solution derived by changing some starting solution to preserve feasibility (i.e., ensure that all the constraints remain satisfied). For instance, a starting solution can be constructed as a feasible solution represented combinatorially by $M-1$ identity permutations $(1,2,\ldots,n)^{T}$. Another example of a starting solution is an ordered set of $M-1$ randomly generated permutations).

%\begin{definition}
  Let $k$ be dimension and $S=(\pi_2, \ldots,\pi_{M})$ be a solution of an instance of the MAP~(\ref{eq:MAP}) with cost coefficients $C$. An $n\times n$ %$N_1\times N_k$ 
  matrix $C(k, S)$ with elements $C_{(k,S)}[i,j]$ is a \emph{projection} of the solution $S$ onto the multidimensional matrix $C$ without (with  free) dimension $k$ is defined as
%  \begin{description}
%    \item[for $d\neq 1$:] $C(d,S)=(c[i ,\pi_1(i), \ldots,\pi_{d-1}(i), j,\pi_{d+1}(i), \ldots, \pi_{D-1}(i)])_{i,j=1,\ldots,N}$, where $j$ is in dimension $d$;
%    \item[for $d=0$:] $C(d,S)=(c_{i \pi_1(j) \ldots  \pi_{D-1}(j)})_{i,j=1,\ldots,N}$, where $j$ is in dimensions $1,2,\ldots,D-1$.
%  \end{description}
\begin{eqnarray}\label{eq:alla.proj}
% \nonumber to remove numbering (before each equation)
  C_{(k,S)}[i,j] = 
  C[i ,\pi_2(i), \ldots,\pi_{k-1}(i), j,\pi_{k+1}(i), \ldots, \pi_{M}(i)], \\ %\nonumber \textrm{ for }i,j=1,\ldots,N, \\
  \label{eq:alla.proj1}
  C_{(1,S)}[i,j] = C[i, \pi_2(j), \ldots,  \pi_{M}(j)], \\ \nonumber
  \textrm{ for } i=1,\ldots,n;j=1,\ldots,n. %i=1,\ldots,N_1;j=1,\ldots,N_k.
\end{eqnarray}
%\end{definition}

Consider the linear assignment problem (LAP) with an $n\times n$ %$N_i\times N_j$ 
matrix of cost coefficients $B=(b_{ij})_{i=1,\ldots,n;j=1,\ldots,n}$: %$B=(b_{ij})_{i=1,\ldots,R;j=1,\ldots,C}$:
\begin{equation}\label{eq:LAP}
  \min_{\pi \in \Pi_n} \sum_{i=1}^{n} b_{i \pi(i)}. %\min_{\pi \in \Pi_N} \sum_{i=1}^{N} b_{i \pi(i)}.
\end{equation}
Let $\textrm{LAP}(B) = \left(\sigma,y(\sigma)\right)$ denote a solution $\sigma \in \Pi_n$ %$\sigma \in \Pi_N$ 
with objective value $y=y(\sigma)$ of the LAP~(\ref{eq:LAP}) with coefficient matrix $B$.

This algorithm works by using a multi-start for exploration and solving the respective two-dimensional assignments for exploitation of a very large neighborhood. First by finding a projection of the multidimensional costs in dimension $k$, the algorithm computes the $n\times n$  %$N_1\times N_k$ 
cost matrix $C(k,S)$ to be used in solving the linear assignment problem (LAP). Then the solution of the LAP replaces the $k$-th %$d$-th 
permutation-column in the current solution, and the algorithm continues in the next dimension $k+1$ for $k<M$ or in the dimension $1$ for $k=M$, until no improvement is possible. %$d+1$ for $d<D$ or in the dimension $1$ for $d=D$, until no improvement is possible.

\begin{figure}[h]
\centering
\includegraphics[height=1.4in, width=1.19in]{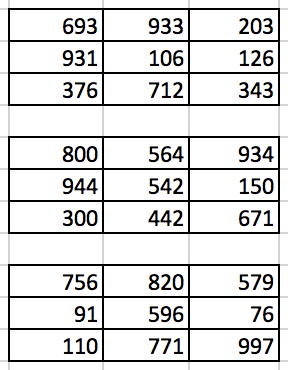}%[height=2in, width=1.7in]{matr}
\caption{\label{fig:matr} Three dimensional input hypermatrix example for algorithm \ref{alg:VLSN.OP}}
\end{figure}

\begin{figure}[h]
\centering
\includegraphics[height=2.25in, width=2.25in]{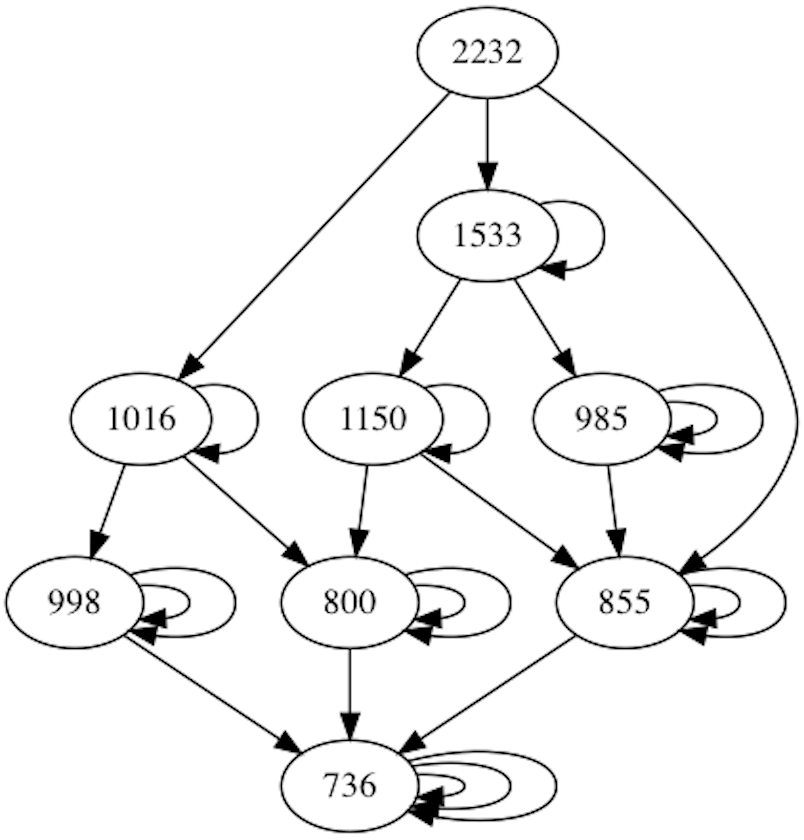}%[height=3in, width=3in]{VLSN}
\caption{\label{fig:vlsn} VLSN for matrix at figure \ref{fig:matr}. Starting solution is (693,542,997) = 2232; result is (203, 442, 91) = 736}
\end{figure}

\begin{algorithm}[h!]
\caption{Very Large-Scale Search via Optimal Permutation}%{Basic Local Search}
\label{alg:VLSN.OP}
\begin{algorithmic}[1]
\STATE Input $\mu$, $M$, $n$, %Initialize 
$C = \{ c_{i_1 \ldots i_M} | i_k=1,\ldots,n, %N_k%, 
k=1,\ldots,M \}$
\STATE Input or Generate $\mu$ starting solutions 
$\Sigma= \{S_1,\ldots,S_{\mu} \}$ %Select $\Sigma$ with $\mu=|\Sigma|$ %$M=|\Sigma|$ %a multi-start strategy $S=\{s_1,s_2,\ldots,s_M\}$ with $M$ (feasible) initial solutions
\FORALL{$S=(s_2,\ldots,s_{M})\in \Sigma$}%\FORALL{$S=(s_2,\ldots,s_{D})\in \Sigma$}%%%{ initial solution $s = (\pi_1,\pi_2,\ldots,\pi_{D-1})$ in $S$ with objective value $y$}
\STATE Compute objective value $y\leftarrow y(S)$ via Equation~(\ref{eq:MAP})
\REPEAT
    \FOR{ dimension $k=1$ \TO $M$} %\FOR{ dimension $d=1$ \TO $D$} %$k=1,2,\ldots,d-1$
    \STATE Get projection $C(k,S)$ %$C(d,S)$ 
    of multidimensional matrix $C$ %Get $\textrm{LAP}\left(C(d,S)\right)$ %Get the LAP in the $d$-th dimension by fixing all $\pi_j$ with $j\neq d$ in %    %    %    Eq.%~(\ref{eq:MAP})
    \STATE Solve $\textrm{LAP}\left(C(k,S)\right) = (\pi_k^{\ast},y_k^{\ast})$ %Solve $\textrm{LAP}\left(C(d,S)\right) = (\pi_d^{\ast},y_d^{\ast})$ %Solve this LAP (e.g., via Hungarian algorithm) to find new $\pi_d^{\ast}$ and the %    %    respective objective value $y_k^{\ast}$
    \ENDFOR
    \STATE Find $k^{\prime}$ such that $y_{k^{\prime}} = \min\{y_k^{\ast}: k=1,\ldots,M\}$ %\STATE Find $d^{\prime}$ such that $y_{d^{\prime}} = \min\{y_d^{\ast}: d=1,\ldots,D\}$ %%%$y_{d'} = \max_{d} y_d^{\ast}$
    \IF{$y_{k^{\prime}} < y$} %\IF{$y_{d^{\prime}} < y$}
        \STATE Update %Move to new starting solution
        $S \leftarrow (s_2,\ldots,s_{k^{\prime}-1},\pi_{k^{\prime}}, s_{k^{\prime}+1}, \ldots, s_{M})$ %%%$S \leftarrow (s_2,\ldots,s_{d^{\prime}-1},\pi_{d^{\prime}}, s_{d^{\prime}+1}, \ldots, s_{D})$%$S \leftarrow (\pi_1,\pi_2,\ldots,\pi_{d^{\prime}},\ldots,\pi_{D-1})$ %with new $\pi_{d'}$
        \STATE $y \leftarrow y_{k^{\prime}}$
    \ENDIF
\UNTIL{no improving move is left}
\ENDFOR
\STATE Output a feasible solution $S$ of the MAP~\eqref{eq:MAP} and cost $y$
\end{algorithmic}
\end{algorithm}
Example of input problem instance for algorithm \ref{alg:VLSN.OP} is depicted at figure \ref{fig:matr}, solution for this problem instance is depicted at figure \ref{fig:vlsn}. Nodes of the graph at figure \ref{fig:vlsn} contain objective function value, directed edges are constructed by solving LAP. 

Figure \ref{fig:vlsn} shows example output graph generated by algorithm \ref{alg:VLSN.OP} with input matrix as in figure \ref{fig:matr}.

In Algorithm~\ref{alg:VLSN.OP}, the steepest descent through dimensions is described. Besides the steepest descent, the following variants of this search can be considered. 
\begin{itemize}
    \item Best improvement-based solution update (place a taboo on checking dimension $k^{\prime}$ %$d^{\prime}$ 
    in the next step).
    \item First improvement-based solution update (continue in new directions by increasing dimensions $k=k^{\prime}+1,\ldots$ %$d=d^{\prime}+1,\ldots$ 
    or revisit previous directions by decreasing $k=k^{\prime}-1,\ldots$, %$d=d^{\prime}-1,\ldots$, 
    placing a taboo on checking dimension $k^{\prime}$ %$d^{\prime}$ 
    in the next step).
    \item Random choice among improving solutions (with taboo on dimension $k^{\prime}$ %$d^{\prime}$ 
    in the next iteration).
\end{itemize}

\subsection{Multi-start strategies for the VLSN search}

The exploration stage of the VLSNS metaheuristic includes construction of another feasible solution from which  the local search is to be restated. Multi-start strategy is a systematic way of constructing starting solutions. The original paper~\cite{kammerdiner2017very} that introduces the VLSNS metaheuristic describes two types of multi-start strategies, namely a random sampling strategy and a design based strategy. 
Using design based multi-start strategy in~\cite{kammerdiner2017very}, we propose a grid based strategy and compare the grid-based multi-start to random sampling multi-start. In particular, we modify the (random) design based restarts into the grid based restarts by using an identity permutation in place of a randomly generated permutation in the construction of starting solutions.

In combinatorics, a cyclic permutation is typically defined as a permutation which shifts all its elements $\{1,\ldots,N\}$ by a fixed offset, so that the elements moved off the end are inserted at the beginning of the permutation.  In~\cite{kammerdiner2017very}, authors consider two types of cyclical permutations applied on a given permutation $\pi = (\pi(1),\ldots,\pi(N))$ of $N$-element set, namely 
\begin{enumerate}
    \item[(i)] a $k$-push-up $\nu_k$:
\begin{eqnarray}\label{eq:pushup}
 \nu_k(\pi(i)) = \left\{
                \begin{array}{rl}
                    \pi(i+k) & \textrm{for }i = 1,\ldots,N-k,  \\
                    \pi(i+k-N) & \textrm{for }i = N-k+1,\ldots,N.
                \end{array}
                \right.
\end{eqnarray}
    \item[(ii)] a $k$-push-down $\delta_k$:
\begin{eqnarray}\label{eq:pushdown}
 \delta_k(\pi(i)) = \left\{
                \begin{array}{rl}
                    \pi(N+i-k) & \textrm{for }i = 1,\ldots,k,  \\
                    \pi(i-k) & \textrm{for }i = k+1,\ldots,N.
                \end{array}
                \right.
\end{eqnarray}
\end{enumerate}
where $k$ is an integer $k = 1,\ldots,N-1$.
%Wolfram DEFINITION: A permutation which shifts all elements of a set by a fixed offset, with the elements shifted off the end inserted back at the beginning. For a set with elements a_0, a_1, ..., a_(n-1), a cyclic permutation of one place to the left would yield a_1, ..., a_(n-1), a_0, and a cyclic permutation of one place to the right would yield a_(n-1), a_0, a_1, ....
%
% mapping can be written as a_i->a_(i+k (mod n)) for a shift of k places. A shift of k places to the left is implemented in the Wolfram Language as RotateLeft[list, k], while a shift of k places to the right is implemented as RotateRight[list, k].

\begin{algorithm}[h!]
\caption{Grid Multi-Start for Very Large-Scale Search}%{Basic Local Search}
\label{alg:MS.grid}
\begin{algorithmic}[1]
\STATE Input $M$, $N$, and an initial permutation $\pi$ (e.g., $\pi = \iota = (1,2,\ldots,N)$)
\STATE Initialize permutation subset to $\Pi_0 \leftarrow \{ \pi \}$
    \FOR{ position offset $k=1$ \TO $N-1$} %\FOR{ dimension $d=1$ \TO $D$} %$k=1,2,\ldots,d-1$
    \STATE Construct $k$-push-up $\nu_k \leftarrow \nu_k(\pi)$ 
    \STATE Add %$k$-push-up 
    $\nu_k$ to $\Pi_0$: $\Pi_0 \leftarrow \Pi_0 \cup \{ \nu_k \}$
    \ENDFOR
\STATE Initialize multi-start (feasible) solutions set $\Sigma \leftarrow \varnothing$
\FOR{number of starting solutions $\mu = 1$ to $N^{M-1}$} %\FOR{ dimension $m=2$ \TO $M$} 
    \STATE $S_{\mu} \leftarrow (\pi,\pi_2,\ldots,\pi_M)$, where $\pi_j \in \Pi_0$ for $j=2,\ldots,M$
    \STATE Add $S_{\mu}$ to $\Sigma$: $\Sigma \leftarrow \Sigma \cup \{ S_{\mu} \}$
\ENDFOR
\STATE Output a multi-start strategy $\Sigma = \{S_1,\ldots,S_{\mu}\}$ with $\mu = N^{M-1}$
\end{algorithmic}
\end{algorithm}

{\color{blue} %edede
\section{Complexity}\label{sec:complexity}
The MAP is NP-hard (Karp, 1972).  Next, we discuss an approach to evaluating the complexity of the VLSN search. In particular, we show that the search tree admits a partition of its nodes into the levels. %, where these 
These levels are ordered based on the MAP dimensions, which can be searched further. We use the partition of nodes into the levels in the search tree to define two related absorbing Markov chains. These two Markov chains allow us to obtain the upper bound on the expected number of moves through partition levels until reaching a local optimum.  

To evaluate complexity of the VLSN search algorithm, observe that this algorithm works by successively searching through dimensions $1, 2, \ldots, M$. Let $y$ and $y^{\prime}$ denote objective values of the current solution $S$ and the neighboring solution $S^{\prime}$. A starting solution or the current solution $S$ has some $I \leq M$ dimensions $m_1, m_2, \ldots, m_I$ that lead to improved objective value $y^{\prime}< y$. For the remaining $M-I \geq 0$ dimensions, there is no improvement as the neighborhood search returns $y^{\prime} = y$. Let us refer to the former $I$ dimensions as the \textit{improving dimensions} of the current solution $S$, while the latter $M- I$ dimensions are called \textit{non-improving dimensions}. The numbers of improving and non-improving dimensions $I, M-I \in \{0,1,\ldots,M\}$. This introduces a natural partition of the search tree of this algorithm into $M+1$ levels according to the specific number $I$ of improving dimensions that remain. 

To illustrate this partition, consider, for example, a search tree in Fig.~\ref{fig:vlsn} for the MAP with $M=3, n=3$. The source node depicts a starting solution with objective value $2232$. The source node has an outdegree of three that corresponds to three improving dimensions, i.e., the source node belongs to the $I =3$ level. The intermediate nodes with objective values $1533, 1150, 1016$ have outdegree two each (not counting the self-loops) and belong to the $I = 2$ level. Whereas the intermediate nodes with objective values $998, 985, 855, 800$ have outdegree one each and belong to the $I = 1$ level. The sink node with the (local and global) minimum objective value $736$ has a zero outdegree and belongs to the $I = 0$ level in the search tree. Observe that the $I = 2$ level's node with value $1533$ %can 
leads to either another node (value $1150$) on the same $I = 2$ level or the node (value $985$) on the lower level $I = 1 (=2 -1)$. Interestingly, the source node (value $2232$) cannot stay on the same top level $I =3$ and instead leads to either the $I = 2$ level nodes or the $I = 1$ level node (value $855$). That is because an improving dimension that is used cannot be replaced with any other previously improved-on dimension.

As the algorithm progresses, the current solution moves from $S$ with $y$ and $I$ to its neighboring solution $S'$ with $y'< y$ and $I^{\prime}$. It is easy to see that for the VLSN search algorithm, we have:
\begin{enumerate}
    \item [(a)] if $I = M$ then $I^{\prime} \in \{I-1,I -2,\ldots,0 \}$;
    \item [(b)] if $I = 1, \ldots, M-1$ then $I^{\prime} \in \{ I, I -1,\ldots,0\}$; 
    \item [(c)] if $I = 0$ then $I^{\prime} = \emptyset$.
\end{enumerate}
 In fact, $S$ and $S^{\prime}$ only differ in one dimension $k^{\ast}$, and by construction of the VLSN search no further improvements can be made through that dimension from $S^{\prime}$. If the current solution $S$ has $I= M$ then this solution is on the top level of the search tree. So, there are no dimensions that have been explored before such $S$. That means that its neighborhood solution $S^{\prime}$ cannot remain on the top level. Instead such $S^{\prime}$ must have $I^{\prime} < I$. On the other hand, if the current solution $S$ has $0<I<M$ then its neighborhood solution $S^{\prime}$ could have an improvement in some previously explored dimension $k_{1}$ (through which $S^{\prime}$ would have a neighbor $S_1$ with value $y_1$) as long as $y_1 < y^{\prime}$. This means that for the levels $I = 1, \ldots,M-1$, it is possible for $S^{\prime}$ to remain at the same level $I$ as $S$ after an objective-value improving move. Obviously, if the current solution $S$ has $I=0$ then no dimensions with improving moves remain, and so the current solution is a local (or global) optimum.
 
 Next, we define two related Markov chains that allow us to evaluate the expected number of transitions through intermediate levels before reaching a minimum. %Any minimum is at an $I = 0$ level, which is an absorbing state. 
 The first Markov chain corresponds with the VLSN search through the levels, with the $I = 0$ level as an absorbing state. The second Markov chain simplifies the evaluation of the first passage times in comparison with the first Markov chain.  
 
 For any $I_0,I_1 = 0,1,\ldots,M$, let $p(I_0,I_1) =\Pr (I_0^{\prime} = I_1)$ denote the probability of the number of improving dimensions changing from $I_0$ to $I_1$ when the VLSN search moves from $S_0$ and $S_1$. 
 %For any $I \in \{M-1, \ldots, 1\}$, let $p(I,I) = \Pr (I^{\prime} = I)$ denote the probability of the number of improving dimensions $I$ remaining the same $I^{\prime} =I$ as the search moves from $S$ to $S^{\prime}$. Let $p(I,I-1) =\Pr (I^{\prime} = I-1) = 1 - p(I,I)$ denote the probability of the number of improving dimensions reducing to $I - 1$, as no more improving moves can be made in dimension $m = k^{\prime}$. Let $p(M,M-1)= \Pr (M^{\prime} = M-1)$ and $p(M,M-2)= \Pr (M^{\prime} = M-2) = 1 - \Pr (M^{\prime} = M-1)$ denote the probabilities of the number of improving dimensions reducing to $M-1$ and $M-2$. 
%
Now let us define a Markov chain $C$ with $M+1$ states corresponding to the levels of the search tree with a given number $I$ of the remaining improving dimensions. This Markov chain has an absorbing state at $I = 0$, as the VLSN search algorithm stops when a local optimum is reached and no improving moves can be made. Consider the transition matrix $P = (P_{ij})_{i,j = 0}^{M}$ of this Markov chain $C$. %Matrix $P$ is an upper bi-diagonal $(M+1)\times(M+1)$ matrix with $P_{I I} = p(I,I)$ and $P_{I I-1 } = 1 - p(I,I)$ and $P_{I j}=  0$ for any $j \neq I,I-1$.
\begin{equation}
    P 
    %= \left(
    %\begin{array}{cccc}
    %    p(0,0) & p(0,0) &\ldots & p(0,M)\\
    %    p(1,0) & p(1,1) &\ldots & p(1,M)\\
    %    \vdots & & & \vdots\\
    %    p(M,0) & p(M,1) & \ldots & p(M,M)
    %\end{array}
    %\right)
    = \left(
    \begin{array}{cccccc}
        1 & 0 & 0 &\ldots & 0 &0\\
        p(1,0) & p(1,1) & 0      &\ldots & 0 &0\\
        \vdots &\vdots  & \vdots & \ddots & \vdots & \vdots\\
        p(M-1,0) & p(M-1,1) & & \ldots & p(M-1,M-1) &0\\
        p(M,0) & p(M,1) & p(M,2) & \ldots & p(M,M-1) &0
    \end{array}
    \right)
\end{equation}
To get an upper bounds for the expected number of transitions before reaching the absorbing state for Markov chain $C$, consider a related Markov chain $C_1$ whose transition matrix $P_1$ is based on the transition probabilities.  
\begin{equation}
    P_1 
    %= \left(
    %\begin{array}{cccc}
    %    p(0,0) & p(0,0) &\ldots & p(0,M)\\
    %    p(1,0) & p(1,1) &\ldots & p(1,M)\\
    %    \vdots & & & \vdots\\
    %    p(M,0) & p(M,1) & \ldots & p(M,M)
    %\end{array}
    %\right)
    = \left(
    \begin{array}{ccccccc}
        1         & 0     & 0 &\ldots & 0 & 0 &0\\
        1 -p(1,1) & p(1,1) & 0  &\ldots & 0& 0 &0\\
        \vdots  &\vdots & \vdots & \ddots & \vdots & \vdots & \vdots\\
        0       & 0      & 0 & \ldots & 1 - p(M-1,M-1)& p(M-1,M-1) &0\\
        0 & 0 & 0 & \ldots &0 & 1 &0
    \end{array}
    \right)
\end{equation}

Let $\nu_I$, $I = 0,1,\ldots, M$ denote the expected number of transitions in this Markov chain $C_1$ starting in state $I$ before reaching the absorbing state $I = 0$, i.e. local minimum. Note that $\nu_{0} = 0$ and for any $0<I < M$, $\nu_I = 1 + p_{I,I}\nu_I + (1 - p_{I,I})\nu_{I-1}$ or $\nu_I = \nu_{I-1} + \frac{1}{1-p_{I,I}} $. Hence, $\nu_I = \sum_{i=1}^{I}\frac{1}{1 - p_{i,i}}$ for $I = 1,2,\ldots,M-1$. Also, $\nu_M = 1 + \nu_{M-1} = 1+ \sum_{i=1}^{M-1}\frac{1}{1-p_{i,i}}$.

The Hungarian algorithm for solving the LAP in each dimension has complexity of order $n^3$. Whereas the complexity of extracting the matrix of cost coefficients for the LAP (i.e., getting the projection $C(k,S)$ of multidimensional cost array $C$) has order $n^2$. The %If the 
expected number of times the VLSN search repeatedly solves the Hungarian algorithm does not exceed %is at most the product 
$\mu M \nu_M = \mu M \left( 1+ \sum_{i=1}^{M-1}\frac{1}{1-p_{i,i}} \right)$. 
Hence, the VLSN search has the  worst-case complexity of $n^3 \mu M \nu_M$.
}%%%%%%%%%%%%%%%%%%%%%%% END COLOR BLUE

\section{Numerical experiments}\label{sec:exp}
The main goal of our numerical studies is to document and compare the performance of two recent heuristics developed for general MAPs of large size. Additionally, we want to understand the extent to which different exploration or multi-start strategies can influence the solutions returned by the VLSN search algorithm. This is particularly important because~\cite{kammerdiner2017very} proposed two distinct %alternative
multi-start strategies but have not studied or compared them numerically. 

Specifically, we have implemented two algorithms described above, and performed  experiments, aimed at answering the following questions:
\begin{enumerate}
  \item Does the VLSN search result in better MAP solutions than those produced by Greedy algorithm?
  \item Does a multi-start VLSN search strategy improve the found solution when compared to a single-start?
    \item Does the use of greedy algorithm's solution as the starting solution for the VLSN search improve the found solution when compared to the VLSN search started at a randomly constructed feasible solution?
    %Can improvement of the VLSN search be achieved  for the search to start it from the solution made by greedy algorithm?
  %\item What are characteristics of recall for tripartite  user profile matching (i.e., finding correspondence of records from three data sources)?
  \item How does the grid compare to the random start?
\end{enumerate}

The algorithms were implemented in C++; the code was executed in sequential manner using 40 core Intel(R) Xeon(R) CPU E5-2640 v4 @ 2.40GHz processor with 256 GB RAM. Optimal solutions were obtained using Gurobi 9.0 solver on the same computer, using 20 threads. In Gurobi, we have set timelimit 10 minutes for each of the problems. While running VLSN we have set limit on number of explored nodes for instances with M = 3 and N = 50 (250000 nodes), and N = 100 (400000 nodes), and with M = 4, N = 20 (1600000 nodes), and N = 30 (2700000 nodes). 

{\color{blue} We have implemented two versions of genetic algorithm, inspired by a random-key generic algorithm as described in \cite{10.1007/s10898-013-0105-7}. As MAP is a constrained discrete optimization problem, we had to develop an encoding and decoding scheme that would satisfy problem constraints  after the crossover operations. We have represented a chromosome as a matrix $M_{cd}$ where number of rows $c$ equals to problem cardinality, and number of columns $d$ to dimensionality of considered MAP. Then, each row of the matrix would represent coordinates of the cell of the MAP hypermatrix. Each column $i \in (1, ... d)$ of $M_{cd}$ represents a permutation of elements $(1, ... , c)$. MAP objective function value would then be equal to the sum of values corresponding to all cells in $M_{cd}$. Crossover operation producing one chromosome from two was implemented as a random combination of columns of two matrices $M^1_{cd}$  and $M^2_{cd}$. Each evolution step was taking $p\%$ of the best (``elitist") solutions, performing crossover resulting in $r\%$ of new chromosomes, merging the results with  $q < p$\% top solutions and generating remaining population randomly.
We have implemented two versions of genetic algorithm: GA1 had $p$ set to 60\%, $q$ to 20\%, and $r$ to 60\%, and GA2 had $p$ set to 40\% (thus, only ``elite" 20\% and remaining 20\% were involved in the crossover). GA1 had 2000 evolution steps, GA2 had 5000 steps.

}

We have generated random %synthetic 
hypermatrices of different dimensions $M$ and cardinalities $N$, and solved the MAP for these synthetic matrices. Matrix elements contain uniformly distributed random numbers in the range from 0 to 1. Results of the experiment are shown in Tables ~\ref{exp-3AP} and \ref{exp-4AP}. 
%{\textcolor{green}{These tables show the average gaps for several versions of the VLSN search and greedy algorithms.}
\textcolor{blue}{These tables show pairwise comparisons of  several versions of the VLSN search with greedy algorithms, genetic algorithms, and all-purpose solver Gurobi~9.
 These include the VLSN search with a single starting solution (denoted by \textbf{VLSN1}), the VLSN search with multiple starts (denoted by \textbf{VLSNMS}), a greedy algorithm (denoted by \textbf{Greedy}), the VLSN search which uses deterministic grid-based strategy to construct starting feasible solutions (denoted by \textbf{Grid-VLSN}), and a hybridization of the VLSN search with greedy algorithm, where the VLSN search uses the solution found by greedy procedure as a starting solution. For each of these algorithms, we record the (best or smallest local optimal) values $y_{\textrm{alg}}^i$ of the solutions found by the algorithm for a given problem instance $i\in \{1,\ldots,I\}$and compare these values using statistical approaches.} %  to the respective values $y_{\textrm{opt}}^i$ of the global optimal solutions computed in Gurobi~9. We compute the average gap}
%\begin{equation}
%    \label{eq:avg.gap}
%    \textrm{average gap} = \frac{1}{I}\sum_{i=1}^{I}\left\{\frac{y_{\textrm{alg}}^i - y_{\textrm{opt}}^i}{y_{\textrm{opt}}^i}\right\}.
%\end{equation}

%{\textcolor{green}{As shown in Table~\ref{exp-2}, on small MAPs, the random multi-start VLSN search and the grid-based multi-start VLSN search algorithms perform the best and are able to find the global optimum. For all tested MAPs, pure greedy approach significantly under-performs in terms of the average gap~\eqref{eq:avg.gap} to the global solution compared with any VLSN search, including hybridization of VLSN with greedy algorithm. As the cardinality and the size of instances grow, the random multi-start VLSN search seems to perform the best among the tested heuristics. However, the deterministic grid-based multi-start begin to outperform the random multi-start search as the cardinality becomes high enough (for example, the 3-APs with $N=100$ and the 4-APs with $N=20$. This suggests that due to the space-filling property of this design-based multi-start~\cite{kammerdiner2017very}, the grid-based multi-start strategy tends to result in exploration of more diverse or expansive set of solutions. On the other hand, the use of Greedy solution as a start in the VLSN search may improve the performance. In fact, for the 3-APs with $N=50$ and $N=100$, hybrid Greedy-VLSN search significantly outperforms other tested heuristics.   } 

{\color{blue}We present and discuss the results of statistical comparison of VLSN and non-VLSN algorithms in Section~\ref{sec:stat-exp}. As the size $N^M$ of hypermatrix of cost coefficients grows exponentially with the dimensionality $M$, we limit our numerical experiments to the MAP instances with $M=3,4$. For the 3-AP, we consider instances with cardinality $N = 3, 10, 20, 30, 40, 50, 100$. While for the 4-AP, we consider $N=3, 10, 20, 30$. Although the MAP instances with $N=3$ are trivial having the input size of 27 or 81 cost coefficients, the larger cardinality $N$ values in our experiments produce hypermatrices with 1,000,000 and 810,000 coefficients and instances with $100!^2$ and $30!^3$ feasible solutions for this generally NP-hard problem.

Furthermore, we illustrate the application of our MAP-based approach to solving multipartite entity resolution problem from computer science in Section~\ref{sec:real-exp}. We show that VLSN-based approaches such as Greedy-VLSN can perform well when matching data from three datasets or sources with various degree of dissimilarity in the records contained in different sources (i.e., datasets).
}

%%%%%%%%%%%%%%%%%%%%%%%%%%% START COLOR BLUE 
{\color{blue} %edede
\subsection{Statistical comparison of the solution quality for the VLSN search and alternative algorithms}\label{sec:stat-exp}

We compare the quality of solutions found by various versions of the VLSN search and alternative algorithms, including metaheuristics, Greedy heuristic, and a standard solver such as Gurobi. In particular, the comparisons are performed on sets of randomly generated problem instances for given cardinality ($n$) and dimensionality ($M$) parameters of the MAP. %other metaheuristic algorithms. 
To perform proper statistical comparisons among the algorithms, we fix the MAP parameters $n,M$ and choose pairs $(A_1,A_2)$ of algorithms to compare so that Algorithm~1, $A_1$, is a version of the VLSN search (e.g., a grid multi-start VLSN search, or a hybrid VLSN search with a start at Greedy solution), and Algorithm~2, $A_2$ is an alternative, non-VLSN algorithm (e.g., a Genetic algorithm GA, a Greedy algorithm, Gurobi, etc.).

Let $y_1, y_2$ (and $T_1,T_2$) denote the objective values (and the number of solutions explored) that are found by algorithms $A_1,A_2$. Recall that $A_1$ always stands for some VLSN algorithm while $A_2$ is non-VLSN, and consider a difference $y_1 - y_2$ between their returned objective values (alternatively, $T_1 - T_2$ is a difference in the number of searched solutions between $A_1$ and $A_2$). This difference~$y_1 - y_2$ is positive, when the quality of solution found by the VLSN search is better than its alternative's $A_2$. For some of $I = 100$ randomly generated problem instances, the~$y_1 - y_2$ differences are randomly distributed. Recall that the objective values $y_1,y_2$ are sums of randomly generated cost coefficients. So, by the law of large numbers, for large cardinality $n$, the difference $y_1 - y_2$ is approximately normally distributed as a difference of approximately normal $y_i, i=1,2$. 

To statistically compare the solution qualities, we construct a confidence interval for $y_1 - y_2$, where the observed differences are taken between $y_1 = y_1(i)$ and $y_2 = y_2(i)$ when $A_1,A_2$ are ran on the \textit{same} problem instance $\mathrm{MAP}_i(n,M)$ where $n,M$ are the cardinality and the dimensionality, and $i = 1,2,\ldots, I$.  To construct the confidence intervals, find the sample mean $$\color{red} A_{1,2} =\widehat{y_2 - y_1} = \frac{1}{I}\sum_{i = 1}^{I} \left(y_2(i) - y_1(i) \right),$$ 
and the sample standard deviation $\color{red} D_{1,2} = \sqrt{\frac{1}{I-1} \sum_{i = 1}^{I} \left(y_2(i) - y_1(i) - A_{1,2} \right)^2 }.$ Given a p-value of $0.05$, %or a 95\% confidence level, 
the %$2-\sigma$ 
95\%-confidence interval for the difference $y_2 - y_1$ is $$\color{red}\left( A_{1,2} - t_{1-\alpha/2,I-1}\cdot \frac{D_{1,2}}{\sqrt{I-1}}, \quad A_{1,2} + t_{1-\alpha/2,I-1}\cdot \frac{D_{1,2}}{\sqrt{I-1}} \right),$$
where $t_\alpha$ is the $\alpha$-th percentile of the %standard normal 
t-distribution. Here, we use $\alpha = 0.05$, $I-1 = 99$, and $t_{1-\alpha/2,I-1} = t_{0.975,99} = 1.984$. %$Z_{1-\alpha/2} = Z_{0.975} = 1.96$.
%%%%%%%%%%%%%%%%%%%%%%%%%%%%%%%%%%%%%%%%%%%%%%%%%%%%%%%%%%%%%%%%%%%%%%%%%%%%%%%%%%% 
%% https://www.itl.nist.gov/div898/software/dataplot/refman1/auxillar/conflimi.htm
%%%%%%%%%%%%%%%%%%%%%%%%%%%%%%%%%%%%%%%%%%%%%%%%%%%%%%%%%%%%%%%%%%%%%%%%%%%%%%%%%%%

We compute $A_{1,2}$, $D_{1,2}$, and the confidence intervals for all possible combinations of the VLSN search variants $A_1$ and the alternative, non-VLSN solution algorithms $A_2$. Using the confidence intervals, we interpret the results of comparison as follows. If zero does not belong to the confidence interval, then the difference in performance (i.e., solution quality) between $A_1$ and $A_2$ is statistically significant. 
%Moreover, the VLSN search outperforms if $0< A_{1,2} - 2 D_{1,2}$, or equivalently $A_{1,2} > 2 D_{1,2}$. Otherwise the non-VLSN algorithm outperforms if $ A_{1,2} + 2 D_{1,2} < 0$, or  $A_{1,2} < - 2 D_{1,2}$.
Moreover, the VLSN search outperforms if $\color{red}0< A_{1,2} - t_{1-\alpha/2,I-1}\cdot \frac{D_{1,2}}{\sqrt{I-1}}$, or equivalently $\color{red} A_{1,2} > t_{1-\alpha/2,I-1}\cdot \frac{D_{1,2}}{\sqrt{I-1}}$. Otherwise the non-VLSN algorithm outperforms if $\color{red} A_{1,2} + t_{1-\alpha/2,I-1}\cdot \frac{D_{1,2}}{\sqrt{I-1}} < 0$, or  $\color{red}A_{1,2} < - t_{1-\alpha/2,I-1}\cdot \frac{D_{1,2}}{\sqrt{I-1}}$.

Observe that most pairs of VLSN and non-VLSN algorithms $A_1,A_2$ are not based on each other. The only exception from this rule is comparison between Greedy-VLSN (a hybridized VLSN with Greedy solutions as its starts) and Greedy (non-VLSN). By construction,  Greedy-VLSN is guaranteed to always do at least as good as Greedy on the same problem instance. So, the differences of these two will follow a log-normal distribution rather than be approximately normally distributed, since there is a dependence. Hence, we modify the construction of the confidence intervals for these two algorithms.

First, for the Greedy-VLSN and Greedy algorithms $A_1,A_2$, we calculate the sample means and standard deviations for the \emph{logged} differences:
$$\color{red} A_{1,2} =\widehat{\log(y_2 - y_1)} = \frac{1}{I}\sum_{i = 1}^{I} \log \left(y_2(i) - y_1(i) \right),$$ 
and %the sample standard deviation 
$$\color{red} D_{1,2} = \sqrt{\frac{1}{I-1} \sum_{i = 1}^{I} \left(\log\left(y_2(i) - y_1(i)\right) - A_{1,2} \right)^2 },$$
where $\log(\cdot)$ represents a natural logarithm.

Using a modified Cox method~\cite{zhou1997confidence} suggested for moderate sample sizes $I = 100\geq 50$, %an $(1-\alpha) = 0.95$ %
a 95\% confidence interval for the log-difference $\log(y_2 - y_1)$ between Greedy-VLSN $A_1$ and Greedy $A_2$ is 
$$\color{red} (\log L, \log U)= \left( A_{1,2} + \frac{D_{1,2}^2}{2}\pm t_{1-\alpha/2} \sqrt{ \frac{D_{1,2}^2}{2} + \frac{D_{1,2}^4}{2(I-1)} }\right),$$
where $t_\alpha$ is the $\alpha$-th percentile of the %standard normal 
t-distribution. Here again, we use $\alpha = 0.05$, $I-1 = 99$, and $t_{1-\alpha/2,I-1} = t_{0.975,99} = 1.984$. %$Z_{1-\alpha/2} = Z_{0.975} = 1.96$.
Then a 95\% confidence interval for the difference $(y_2 - y_1)$ between Greedy-VLSN $A_1$ and Greedy $A_2$ is 
$$\color{red} ( L, U)= \left( \exp\left\{A_{1,2} + \frac{D_{1,2}^2}{2}\pm t_{1-\alpha/2} \sqrt{ \frac{D_{1,2}^2}{2} + \frac{D_{1,2}^4}{2(I-1)} }\right\}\right),$$
%%%%%%%%%%%%%%%%%%%%%%%%%%%%%%%%%%%%%%%%%%%%%%%%%%%%%%%%%%%%%%%%%%%%%%%%%%%%%%%%%%% 
%% https://www.itl.nist.gov/div898/software/dataplot/refman1/auxillar/conflimi.htm
%%%%%%%%%%%%%%%%%%%%%%%%%%%%%%%%%%%%%%%%%%%%%%%%%%%%%%%%%%%%%%%%%%%%%%%%%%%%%%%%%%%

The results of the comparison for the MAP with $M=3$ and $M=4$ are summarized in Tables~\ref{exp-3AP} and \ref{exp-4AP}, respectively. With exception of the small-size MAP with $M=3,n=3$ or $M=4,n=4$,
VLSN algorithms outperforms Greedy and both genetic algorithms. As the cardinality and the size of instances grow, the random multi-start search (VLSNMS) seems to perform the best among the tested heuristics, as evident from the comparison with Gurobi. However, the deterministic grid-based multi-start (Grid-VLSN) begin to outperform the random multi-start search as the cardinality becomes sufficiently high (e.g., the 3-APs with $n = 100$ and the 4-APs with $n = 20,30$.
 This suggests that due to the space-filling property of this design-based multi-start~\cite{kammerdiner2017very}, the grid-based multi-start strategy tends to result in exploration of more diverse or expansive set of solutions. On the other hand, the use of Greedy solution as a start in the VLSN search may improve the performance. In fact, for the 3-APs with $N=50$ and $N=100$, hybrid Greedy-VLSN search significantly outperforms other tested heuristics. 
%!!ADD A TABLE HERE!!
\begin{center}
%\begin{table}[]
%\centering
\begin{longtable}{|p{.13\textwidth}|p{.10\textwidth}|l|l|l|l|l|l|}%{|p{.13\textwidth} | p{.10\textwidth} |p{.015\textwidth}| p{.015\textwidth} | p{.15\textwidth} | p{.15\textwidth} | p{.20\textwidth} | p{.08\textwidth} |} 
%\centering
\caption{Statistical Comparison of Experiments for %the MAPs with %dimensionality 
$M=3$}
\label{exp-3AP}\\ %{exp-1}
%\begin{tabular}{|l|l|l|l|l|l|l|l|}
\hline
 \textbf{$A_1$:} %VLSN}
  & \textbf{$A_2$:} %Non-VLSN} 
 & \multirow{2}{*}{\textbf{M}}   & \multirow{2}{*}{\textbf{n}} & \textbf{Mean} %$A_{1,2}$} 
 & \multirow{2}{*}{\textbf{Std. $D_{1,2}$}} & \multirow{2}{*}{\textbf{95\% C.I.}} & \multirow{2}{*}{\textbf{Winner}} \\ 
 \textbf{VLSN}& \textbf{Non-VLSN} & & & $A_{1,2}$ & & & \\
 \hline
 \endhead
 
 \hline  \multicolumn{8}{|r|}{\textbf{continued on the next page}}\\ \hline
 \endfoot
 
 \hline %\hline
 \endlastfoot
% \textbf{VLSN1} & \textbf{Greedy} & 3  & 3 &  &  & (,) & VLSN \\
% \textbf{VLSN1} & \textbf{GA1} & 3  & 3 &  &  & (,) & VLSN \\
% \textbf{VLSN1} & \textbf{GA2} & 3  & 3 &  &  & (,) & VLSN \\
% \textbf{VLSN1} & \textbf{Gurobi} & 3  & 3 &  &  & (,) & VLSN \\
% \hline
% \textbf{VLSNMS} & \textbf{Greedy} & 3  & 3 &  &  & (,) & VLSN \\
% \textbf{VLSNMS} & \textbf{GA1} & 3  & 3 &  &  & (,) & VLSN \\
% \textbf{VLSNMS} & \textbf{GA2} & 3  & 3 &  &  & (,) & VLSN \\
% \textbf{VLSNMS} & \textbf{Gurobi} & 3  & 3 &  &  & (,) & VLSN \\
% \hline
% \textbf{Grid-VLSN} & \textbf{Greedy} & 3  & 3 &  &  & (,) & VLSN \\
% \textbf{Grid-VLSN} & \textbf{GA1} & 3  & 3 &  &  & (,) & VLSN \\
% \textbf{Grid-VLSN} & \textbf{GA2} & 3  & 3 &  &  & (,) & VLSN \\
% \textbf{Grid-VLSN} & \textbf{Gurobi} & 3  & 3 &  &  & (,) & VLSN \\
% \hline
% \textbf{Greedy-VLSN} & \textbf{Greedy} & 3  & 3 &  &  & (,) & VLSN \\
% \textbf{Greedy-VLSN} & \textbf{GA1} & 3  & 3 &  &  & (,) & VLSN \\
% \textbf{Greedy-VLSN} & \textbf{GA2} & 3  & 3 &  &  & (,) & VLSN \\
% \textbf{Greedy-VLSN} & \textbf{Gurobi} & 3  & 3 &  &  & (,) & VLSN \\
% \hline
\hline
\multirow{4}{*}{\textbf{VLSN1}} & \textbf{Greedy} & 3  & 3 & 214342.030 & 229466.388 & (168586.546, 260097.514) & VLSN \\
 & \textbf{GA1} & 3  & 3 & -7695.080 & 45128.527 & (-16693.686, 1303.526) & None \\
 & \textbf{GA2} & 3  & 3 & -7695.080 & 45128.527 & (-16693.686, 1303.526) & None \\
 & \textbf{Gurobi} & 3  & 3 & -7695.080 & 45128.527 & (-16693.686, 1303.526) & None \\
\hline
\multirow{4}{*}{\textbf{VLSNMS}} & \textbf{Greedy} & 3  & 3 & 223494.530 & 223403.268 & (178948.029, 268041.031) & VLSN \\
 & \textbf{GA1} & 3  & 3 & 1457.420 & 12369.498 & (-1009.052, 3923.892) & None \\
 & \textbf{GA2} & 3  & 3 & 1457.420 & 12369.498 & (-1009.052, 3923.892) & None \\
 & \textbf{Gurobi} & 3  & 3 & 1457.420 & 12369.498 & (-1009.052, 3923.892) & None \\
\hline
\multirow{4}{2mm}{\textbf{Grid-VLSN}} & \textbf{Greedy} & 3  & 3 & 223494.530 & 223403.268 & (178948.029, 268041.031) & VLSN \\
 & \textbf{GA1} & 3  & 3 & 1457.420 & 12369.498 & (-1009.052, 3923.892) & None \\
 & \textbf{GA2} & 3  & 3 & 1457.420 & 12369.498 & (-1009.052, 3923.892) & None \\
 & \textbf{Gurobi} & 3  & 3 & 1457.420 & 12369.498 & (-1009.052, 3923.892) & None \\
\hline
\multirow{4}{2mm}{\textbf{Greedy-VLSN}} & \textbf{Greedy} & 3  & 3 & 209645.460 & 230090.725 & (3225943.686, 810173971774378.625) & VLSN \\
 & \textbf{GA1} & 3  & 3 & -12391.650 & 41144.578 & (-20595.858, -4187.442) & Non-VLSN \\
 & \textbf{GA2} & 3  & 3 & -12391.650 & 41144.578 & (-20595.858, -4187.442) & Non-VLSN \\
 & \textbf{Gurobi} & 3  & 3 & -12391.650 & 41144.578 & (-20595.858, -4187.442) & Non-VLSN \\
\hline
\hline
\multirow{4}{*}{\textbf{VLSN1}} & \textbf{Greedy} & 3  & 10 & 542017.770 & 315482.629 & (479110.691, 604924.849) & VLSN \\
 & \textbf{GA1} & 3  & 10 & 760288.510 & 175283.909 & (725336.986, 795240.034) & VLSN \\
 & \textbf{GA2} & 3  & 10 & 769004.430 & 181455.812 & (732822.231, 805186.629) & VLSN \\
 & \textbf{Gurobi} & 3  & 10 & -128186.000 & 70074.942 & (-142158.909, -114213.091) & Non-VLSN \\
\hline
\multirow{4}{*}{\textbf{VLSNMS}} & \textbf{Greedy} & 3  & 10 & 665923.850 & 313330.992 & (603445.806, 728401.894) & VLSN \\
 & \textbf{GA1} & 3  & 10 & 884194.590 & 154886.611 & (853310.277, 915078.903) & VLSN \\
 & \textbf{GA2} & 3  & 10 & 892910.510 & 167886.376 & (859434.050, 926386.970) & VLSN \\
 & \textbf{Gurobi} & 3  & 10 & -4279.920 & 12186.237 & (-6709.850, -1849.990) & Non-VLSN \\
\hline
\multirow{4}{2mm}{\textbf{Grid-VLSN}} & \textbf{Greedy} & 3  & 10 & 629492.110 & 318919.295 & (565899.761, 693084.459) & VLSN \\
 & \textbf{GA1} & 3  & 10 & 847762.850 & 156755.347 & (816505.912, 879019.788) & VLSN \\
 & \textbf{GA2} & 3  & 10 & 856478.770 & 169972.853 & (822586.268, 890371.272) & VLSN \\
 & \textbf{Gurobi} & 3  & 10 & -40711.660 & 32373.959 & (-47167.011, -34256.309) & Non-VLSN \\
\hline
\multirow{4}{2mm}{\textbf{Greedy-VLSN}} & \textbf{Greedy} & 3  & 10 & 542307.910 & 320121.278 & (156061.777, 4065828346.104) & VLSN \\
 & \textbf{GA1} & 3  & 10 & 760578.650 & 165832.273 & (727511.777, 793645.523) & VLSN \\
 & \textbf{GA2} & 3  & 10 & 769294.570 & 196599.749 & (730092.678, 808496.462) & VLSN \\
 & \textbf{Gurobi} & 3  & 10 & -127895.860 & 87987.696 & (-145440.563, -110351.157) & Non-VLSN \\
\hline
\hline
\multirow{4}{*}{\textbf{VLSN1}} & \textbf{Greedy} & 3  & 20 & 585363.040 & 307288.252 & (524089.915, 646636.165) & VLSN \\
 & \textbf{GA1} & 3  & 20 & 3823013.580 & 300278.765 & (3763138.144, 3882889.016) & VLSN \\
 & \textbf{GA2} & 3  & 20 & 3835849.410 & 295563.710 & (3776914.153, 3894784.667) & VLSN \\
 & \textbf{Gurobi} & 3  & 20 & -236083.590 & 52816.299 & (-246615.134, -225552.046) & Non-VLSN \\
\hline
\multirow{4}{*}{\textbf{VLSNMS}} & \textbf{Greedy} & 3  & 20 & 722671.110 & 303454.217 & (662162.490, 783179.730) & VLSN \\
 & \textbf{GA1} & 3  & 20 & 3960321.650 & 298155.840 & (3900869.524, 4019773.776) & VLSN \\
 & \textbf{GA2} & 3  & 20 & 3973157.480 & 301518.528 & (3913034.836, 4033280.124) & VLSN \\
 & \textbf{Gurobi} & 3  & 20 & -98775.520 & 20701.131 & (-102903.315, -94647.725) & Non-VLSN \\
\hline
\multirow{4}{2mm}{\textbf{Grid-VLSN}} & \textbf{Greedy} & 3  & 20 & 672606.190 & 304057.858 & (611977.204, 733235.176) & VLSN \\
 & \textbf{GA1} & 3  & 20 & 3910256.730 & 293280.628 & (3851776.719, 3968736.741) & VLSN \\
 & \textbf{GA2} & 3  & 20 & 3923092.560 & 306168.609 & (3862042.692, 3984142.428) & VLSN \\
 & \textbf{Gurobi} & 3  & 20 & -148840.440 & 26144.873 & (-154053.715, -143627.165) & Non-VLSN \\
\hline
\multirow{4}{2mm}{\textbf{Greedy-VLSN}} & \textbf{Greedy} & 3  & 20 & 640749.880 & 309907.226 & (65603.813, 18985277.328) & VLSN \\
 & \textbf{GA1} & 3  & 20 & 3878400.420 & 299689.695 & (3818642.444, 3938158.396) & VLSN \\
 & \textbf{GA2} & 3  & 20 & 3891236.250 & 310487.262 & (3829325.244, 3953147.256) & VLSN \\
 & \textbf{Gurobi} & 3  & 20 & -180696.750 & 56969.899 & (-192056.520, -169336.980) & Non-VLSN \\
\hline
\hline
\multirow{4}{*}{\textbf{VLSN1}} & \textbf{Greedy} & 3  & 30 & 547523.710 & 287548.220 & (490186.738, 604860.682) & VLSN \\
 & \textbf{GA1} & 3  & 30 & 7434182.070 & 426287.497 & (7349180.555, 7519183.585) & VLSN \\
 & \textbf{GA2} & 3  & 30 & 7499214.450 & 292365.595 & (7440916.896, 7557512.004) & VLSN \\
 & \textbf{Gurobi} & 3  & 30 & -272285.320 & 32905.022 & (-278846.565, -265724.075) & Non-VLSN \\
\hline
\multirow{4}{*}{\textbf{VLSNMS}} & \textbf{Greedy} & 3  & 30 & 659756.550 & 291005.873 & (601730.124, 717782.976) & VLSN \\
 & \textbf{GA1} & 3  & 30 & 7546414.910 & 428167.560 & (7461038.512, 7631791.308) & VLSN \\
 & \textbf{GA2} & 3  & 30 & 7611447.290 & 290447.548 & (7553532.193, 7669362.387) & VLSN \\
 & \textbf{Gurobi} & 3  & 30 & -160052.480 & 15141.474 & (-163071.682, -157033.278) & Non-VLSN \\
\hline
\multirow{4}{2mm}{\textbf{Grid-VLSN}} & \textbf{Greedy} & 3  & 30 & 619527.740 & 291727.276 & (561357.466, 677698.014) & VLSN \\
 & \textbf{GA1} & 3  & 30 & 7506186.100 & 432202.493 & (7420005.138, 7592367.062) & VLSN \\
 & \textbf{GA2} & 3  & 30 & 7571218.480 & 287537.072 & (7513883.731, 7628553.229) & VLSN \\
 & \textbf{Gurobi} & 3  & 30 & -200281.290 & 21595.991 & (-204587.520, -195975.060) & Non-VLSN \\
\hline
\multirow{4}{2mm}{\textbf{Greedy-VLSN}} & \textbf{Greedy} & 3  & 30 & 623372.170 & 282526.737 & (104743.167, 536568.429) & VLSN \\
 & \textbf{GA1} & 3  & 30 & 7510030.530 & 430923.987 & (7424104.501, 7595956.559) & VLSN \\
 & \textbf{GA2} & 3  & 30 & 7575062.910 & 283967.037 & (7518440.024, 7631685.796) & VLSN \\
 & \textbf{Gurobi} & 3  & 30 & -196436.860 & 45216.148 & (-205452.937, -187420.783) & Non-VLSN \\
\hline
\hline
\multirow{4}{*}{\textbf{VLSN1}} & \textbf{Greedy} & 3  & 40 & 632734.380 & 310776.474 & (570765.706, 694703.054) & VLSN \\
 & \textbf{GA1} & 3  & 40 & 11370163.470 & 450560.750 & (11280321.881, 11460005.059) & VLSN \\
 & \textbf{GA2} & 3  & 40 & 11159395.770 & 412442.699 & (11077154.901, 11241636.639) & VLSN \\
 & \textbf{Gurobi} & 3  & 40 & -301675.170 & 33257.627 & (-308306.724, -295043.616) & Non-VLSN \\
\hline
\multirow{4}{*}{\textbf{VLSNMS}} & \textbf{Greedy} & 3  & 40 & 741892.330 & 308895.953 & (680298.631, 803486.029) & VLSN \\
 & \textbf{GA1} & 3  & 40 & 11479321.420 & 445664.135 & (11390456.213, 11568186.627) & VLSN \\
 & \textbf{GA2} & 3  & 40 & 11268553.720 & 410541.045 & (11186692.040, 11350415.400) & VLSN \\
 & \textbf{Gurobi} & 3  & 40 & -192517.220 & 13239.799 & (-195157.229, -189877.211) & Non-VLSN \\
\hline
\multirow{4}{2mm}{\textbf{Grid-VLSN}} & \textbf{Greedy} & 3  & 40 & 702126.210 & 310744.835 & (640163.844, 764088.576) & VLSN \\
 & \textbf{GA1} & 3  & 40 & 11439555.300 & 446812.411 & (11350461.127, 11528649.473) & VLSN \\
 & \textbf{GA2} & 3  & 40 & 11228787.600 & 412326.091 & (11146569.983, 11311005.217) & VLSN \\
 & \textbf{Gurobi} & 3  & 40 & -232283.340 & 19798.391 & (-236231.129, -228335.551) & Non-VLSN \\
\hline
\multirow{4}{2mm}{\textbf{Greedy-VLSN}} & \textbf{Greedy} & 3  & 40 & 738094.340 & 302753.314 & (79306.429, 4576228.208) & VLSN \\
 & \textbf{GA1} & 3  & 40 & 11475523.430 & 450631.897 & (11385667.654, 11565379.206) & VLSN \\
 & \textbf{GA2} & 3  & 40 & 11264755.730 & 411792.814 & (11182644.448, 11346867.012) & VLSN \\
 & \textbf{Gurobi} & 3  & 40 & -196315.210 & 42157.588 & (-204721.412, -187909.008) & Non-VLSN \\
\hline
\hline
\multirow{4}{*}{\textbf{VLSN1}} & \textbf{Greedy} & 3  & 50 & 532389.470 & 305910.186 & (471391.131, 593387.809) & VLSN \\
 & \textbf{GA1} & 3  & 50 & 15283969.450 & 570429.146 & (15170226.162, 15397712.738) & VLSN \\
 & \textbf{GA2} & 3  & 50 & 15099111.000 & 503005.517 & (14998811.950, 15199410.050) & VLSN \\
 & \textbf{Gurobi} & 3  & 50 & -378638.790 & 39552.538 & (-386525.546, -370752.034) & Non-VLSN \\
\hline
\multirow{4}{*}{\textbf{VLSNMS}} & \textbf{Greedy} & 3  & 50 & 663933.540 & 303157.072 & (603484.171, 724382.909) & VLSN \\
 & \textbf{GA1} & 3  & 50 & 15415513.520 & 568004.680 & (15302253.669, 15528773.371) & VLSN \\
 & \textbf{GA2} & 3  & 50 & 15230655.070 & 504040.517 & (15130149.642, 15331160.498) & VLSN \\
 & \textbf{Gurobi} & 3  & 50 & -247094.720 & 13459.298 & (-249778.497, -244410.943) & Non-VLSN \\
\hline
\multirow{4}{2mm}{\textbf{Grid-VLSN}} & \textbf{Greedy} & 3  & 50 & 656987.910 & 307018.338 & (595768.606, 718207.214) & VLSN \\
 & \textbf{GA1} & 3  & 50 & 15408567.890 & 568944.808 & (15295120.578, 15522015.202) & VLSN \\
 & \textbf{GA2} & 3  & 50 & 15223709.440 & 503299.612 & (15123351.748, 15324067.132) & VLSN \\
 & \textbf{Gurobi} & 3  & 50 & -254040.350 & 14541.767 & (-256939.971, -251140.729) & Non-VLSN \\
\hline
\multirow{4}{2mm}{\textbf{Greedy-VLSN}} & \textbf{Greedy} & 3  & 50 & 717863.920 & 302082.191 & (132413.175, 547247.574) & VLSN \\
 & \textbf{GA1} & 3  & 50 & 15469443.900 & 569746.864 & (15355836.659, 15583051.141) & VLSN \\
 & \textbf{GA2} & 3  & 50 & 15284585.450 & 505483.372 & (15183792.317, 15385378.583) & VLSN \\
 & \textbf{Gurobi} & 3  & 50 & -193164.340 & 32610.082 & (-199666.774, -186661.906) & Non-VLSN \\
\hline
\hline
\multirow{4}{*}{\textbf{VLSN1}} & \textbf{Greedy} & 3  & 100 & 432980.030 & 308760.169 & (371413.406, 494546.654) & VLSN \\
 & \textbf{GA1} & 3  & 100 & 36266744.820 & 641765.795 & (36138777.040, 36394712.600) & VLSN \\
 & \textbf{GA2} & 3  & 100 & 35997235.880 & 725050.818 & (35852661.107, 36141810.653) & VLSN \\
 & \textbf{Gurobi} & 3  & 100 & -446354.930 & 37218.735 & (-453776.327, -438933.533) & Non-VLSN \\
\hline
\multirow{4}{*}{\textbf{VLSNMS}} & \textbf{Greedy} & 3  & 100 & 572240.510 & 312392.330 & (509949.635, 634531.385) & VLSN \\
 & \textbf{GA1} & 3  & 100 & 36406005.300 & 638373.264 & (36278713.989, 36533296.611) & VLSN \\
 & \textbf{GA2} & 3  & 100 & 36136496.360 & 728320.389 & (35991269.637, 36281723.083) & VLSN \\
 & \textbf{Gurobi} & 3  & 100 & -307094.450 & 10890.716 & (-309266.053, -304922.847) & Non-VLSN \\
\hline
\multirow{4}{2mm}{\textbf{Grid-VLSN}} & \textbf{Greedy} & 3  & 100 & 594828.740 & 312548.849 & (532506.655, 657150.825) & VLSN \\
 & \textbf{GA1} & 3  & 100 & 36428593.530 & 638288.430 & (36301319.135, 36555867.925) & VLSN \\
 & \textbf{GA2} & 3  & 100 & 36159084.590 & 727803.960 & (36013960.843, 36304208.337) & VLSN \\
 & \textbf{Gurobi} & 3  & 100 & -284506.220 & 10877.640 & (-286675.216, -282337.224) & Non-VLSN \\
\hline
\multirow{4}{2mm}{\textbf{Greedy-VLSN}} & \textbf{Greedy} & 3  & 100 & 726941.570 & 309360.760 & (134148.649, 551656.943) & VLSN \\
 & \textbf{GA1} & 3  & 100 & 36560706.360 & 634202.488 & (36434246.699, 36687166.021) & VLSN \\
 & \textbf{GA2} & 3  & 100 & 36291197.420 & 726377.431 & (36146358.122, 36436036.718) & VLSN \\
 & \textbf{Gurobi} & 3  & 100 & -152393.390 & 24995.680 & (-157377.516, -147409.264) & Non-VLSN \\
\hline
\hline
%
%\hline
% \hline
%\end{tabular}
%\caption{Statistical Comparison of Experiments}
\end{longtable}
\end{center}
%\end{table}

%%%%%%%%%%%%%%%%%%%%%%%%%%%%%%%%%%%%%%%%%%%%%%%%%%%%%%%%%%%
%%%%%%%%%%%%%%%%%%%%%%%%%%%%%%%%%%%%%%%%%%%%%%%%%%%%%%%%%%%
%\begin{table}[]
%\centering
%\caption{Statistical Comparison of Experiments}
%\label{exp-1}
%\begin{tabular}{|l|l|l|l|l|l|l|l|}
%\hline
% \textbf{$A_1$: VLSN} & \textbf{$A_2$: Non-VLSN} & \textbf{M}   & \textbf{n} & \textbf{Mean $A_{1,2}$} & \textbf{Std. $D_{1,2}$} & \textbf{$2\sigma $ C.I.} & \textbf{Winner} \\ \hline
\begin{center}
%\begin{table}[]
%\centering
\begin{longtable}{|p{.13\textwidth}|p{.10\textwidth}|l|l|l|l|l|l|}%{|p{.13\textwidth} | p{.10\textwidth} |p{.015\textwidth}| p{.015\textwidth} | p{.15\textwidth} | p{.15\textwidth} | p{.20\textwidth} | p{.08\textwidth} |} 
%\centering
\caption{Statistical Comparison of Experiments for %the MAPs with %dimensionality 
$M=4$}
\label{exp-4AP}\\ %{exp-1}
%\begin{tabular}{|l|l|l|l|l|l|l|l|}
\hline
 \textbf{$A_1$:} %VLSN}
  & \textbf{$A_2$:} %Non-VLSN} 
 & \multirow{2}{*}{\textbf{M}}   & \multirow{2}{*}{\textbf{n}} & \textbf{Mean} %$A_{1,2}$} 
 & \multirow{2}{*}{\textbf{Std. $D_{1,2}$}} & \multirow{2}{*}{\textbf{95\% C.I.}} & \multirow{2}{*}{\textbf{Winner}} \\ 
 \textbf{VLSN}& \textbf{Non-VLSN} & & & $A_{1,2}$ & & & \\
 \hline
 \endhead
 
 \hline  \multicolumn{8}{|r|}{\textbf{continued on the next page}}\\ \hline
 \endfoot
 
 \hline %\hline
 \endlastfoot
\multirow{4}{*}{\textbf{VLSN1}} & \textbf{Greedy} & 4  & 4 & 420708.680 & 291983.871 & (362487.241, 478930.119) & VLSN \\
 & \textbf{GA1} & 4  & 4 & -33290.710 & 64040.578 & (-46060.369, -20521.051) & Non-VLSN \\
 & \textbf{GA2} & 4  & 4 & -33290.710 & 64040.578 & (-46060.369, -20521.051) & Non-VLSN \\
 & \textbf{Gurobi} & 4  & 4 & -33290.720 & 64040.574 & (-46060.379, -20521.061) & Non-VLSN \\
\hline
\multirow{4}{*}{\textbf{VLSNMS}} & \textbf{Greedy} & 4  & 4 & 453999.390 & 273321.510 & (399499.217, 508499.563) & VLSN \\
 & \textbf{GA1} & 4  & 4 & 0.000 & 0.000 & (0.000, 0.000) & Tie \\ % None \\
 & \textbf{GA2} & 4  & 4 & 0.000 & 0.000 & (0.000, 0.000) & Tie \\ % None \\
 & \textbf{Gurobi} & 4  & 4 & -0.010 & 0.100 & (-0.030, 0.010) & None \\
\hline
\multirow{4}{2mm}{\textbf{Grid-VLSN}} & \textbf{Greedy} & 4  & 4 & 453999.390 & 273321.510 & (399499.217, 508499.563) & VLSN \\
 & \textbf{GA1} & 4  & 4 & 0.000 & 0.000 & (0.000, 0.000) & Tie \\ % None \\
 & \textbf{GA2} & 4  & 4 & 0.000 & 0.000 & (0.000, 0.000) & Tie \\ % None \\
 & \textbf{Gurobi} & 4  & 4 & -0.010 & 0.100 & (-0.030, 0.010) & None \\
\hline
\multirow{4}{2mm}{\textbf{Greedy-VLSN}} & \textbf{Greedy} & 4  & 4 & 342323.700 & 286661.924 & (6464833.248, 884074728781398.875) & VLSN \\
 & \textbf{GA1} & 4  & 4 & -111675.690 & 118453.291 & (-135295.217, -88056.163) & Non-VLSN \\
 & \textbf{GA2} & 4  & 4 & -111675.690 & 118453.291 & (-135295.217, -88056.163) & Non-VLSN \\
 & \textbf{Gurobi} & 4  & 4 & -111675.700 & 118453.283 & (-135295.226, -88056.174) & Non-VLSN \\
\hline
\hline
\multirow{4}{*}{\textbf{VLSN1}} & \textbf{Greedy} & 4  & 10 & 457756.410 & 303442.832 & (397250.060, 518262.760) & VLSN \\
 & \textbf{GA1} & 4  & 10 & 849064.340 & 157667.201 & (817625.579, 880503.101) & VLSN \\
 & \textbf{GA2} & 4  & 10 & 944836.200 & 122696.744 & (920370.530, 969301.870) & VLSN \\
 & \textbf{Gurobi} & 4  & 10 & -168329.980 & 35154.080 & (-175339.686, -161320.274) & Non-VLSN \\
\hline
\multirow{4}{*}{\textbf{VLSNMS}} & \textbf{Greedy} & 4  & 10 & 580535.270 & 296673.117 & (521378.798, 639691.742) & VLSN \\
 & \textbf{GA1} & 4  & 10 & 971843.200 & 145473.373 & (942835.882, 1000850.518) & VLSN \\
 & \textbf{GA2} & 4  & 10 & 1067615.060 & 119273.555 & (1043831.972, 1091398.148) & VLSN \\
 & \textbf{Gurobi} & 4  & 10 & -45551.120 & 14835.867 & (-48509.385, -42592.855) & Non-VLSN \\
\hline
\multirow{4}{2mm}{\textbf{Grid-VLSN}} & \textbf{Greedy} & 4  & 10 & 567159.630 & 297875.243 & (507763.455, 626555.805) & VLSN \\
 & \textbf{GA1} & 4  & 10 & 958467.560 & 144629.304 & (929628.549, 987306.571) & VLSN \\
 & \textbf{GA2} & 4  & 10 & 1054239.420 & 120087.033 & (1030294.125, 1078184.715) & VLSN \\
 & \textbf{Gurobi} & 4  & 10 & -58926.760 & 15813.094 & (-62079.883, -55773.637) & Non-VLSN \\
\hline
\multirow{4}{2mm}{\textbf{Greedy-VLSN}} & \textbf{Greedy} & 4  & 10 & 471827.540 & 302512.696 & (287030.494, 48785145828.742) & VLSN \\
 & \textbf{GA1} & 4  & 10 & 863135.470 & 154370.048 & (832354.159, 893916.781) & VLSN \\
 & \textbf{GA2} & 4  & 10 & 958907.330 & 136883.720 & (931612.784, 986201.876) & VLSN \\
 & \textbf{Gurobi} & 4  & 10 & -154258.850 & 64071.598 & (-167034.695, -141483.005) & Non-VLSN \\
\hline
\hline
\multirow{4}{*}{\textbf{VLSN1}} & \textbf{Greedy} & 4  & 20 & 342618.340 & 295625.389 & (283670.785, 401565.895) & VLSN \\
 & \textbf{GA1} & 4  & 20 & 3931114.510 & 337716.280 & (3863774.052, 3998454.968) & VLSN \\
 & \textbf{GA2} & 4  & 20 & 3954816.290 & 282396.423 & (3898506.584, 4011125.996) & VLSN \\
 & \textbf{Gurobi} & 4  & 20 & -269337.000 & 32469.602 & (-275811.423, -262862.577) & Non-VLSN \\
\hline
\multirow{4}{*}{\textbf{VLSNMS}} & \textbf{Greedy} & 4  & 20 & 481352.330 & 295531.385 & (422423.519, 540281.141) & VLSN \\
 & \textbf{GA1} & 4  & 20 & 4069848.500 & 333945.221 & (4003259.989, 4136437.011) & VLSN \\
 & \textbf{GA2} & 4  & 20 & 4093550.280 & 282354.181 & (4037248.997, 4149851.563) & VLSN \\
 & \textbf{Gurobi} & 4  & 20 & -130603.010 & 10977.705 & (-132791.959, -128414.061) & Non-VLSN \\
\hline
\multirow{4}{2mm}{\textbf{Grid-VLSN}} & \textbf{Greedy} & 4  & 20 & 483696.310 & 294034.410 & (425065.995, 542326.625) & VLSN \\
 & \textbf{GA1} & 4  & 20 & 4072192.480 & 333364.431 & (4005719.778, 4138665.182) & VLSN \\
 & \textbf{GA2} & 4  & 20 & 4095894.260 & 282618.431 & (4039540.285, 4152248.235) & VLSN \\
 & \textbf{Gurobi} & 4  & 20 & -128259.030 & 10061.525 & (-130265.293, -126252.767) & Non-VLSN \\
\hline
\multirow{4}{2mm}{\textbf{Greedy-VLSN}} & \textbf{Greedy} & 4  & 20 & 468652.290 & 291875.195 & (76908.751, 479624503.370) & VLSN \\
 & \textbf{GA1} & 4  & 20 & 4057148.460 & 333773.486 & (3990594.193, 4123702.727) & VLSN \\
 & \textbf{GA2} & 4  & 20 & 4080850.240 & 286310.078 & (4023760.153, 4137940.327) & VLSN \\
 & \textbf{Gurobi} & 4  & 20 & -143303.050 & 43853.404 & (-152047.397, -134558.703) & Non-VLSN \\
\hline
\hline
\multirow{4}{*}{\textbf{VLSN1}} & \textbf{Greedy} & 4  & 30 & 301879.240 & 299150.600 & (242228.759, 361529.721) & VLSN \\
 & \textbf{GA1} & 4  & 30 & 7477163.470 & 324465.621 & (7412465.187, 7541861.753) & VLSN \\
 & \textbf{GA2} & 4  & 30 & 7437036.990 & 369574.419 & (7363344.035, 7510729.945) & VLSN \\
 & \textbf{Gurobi} & 4  & 30 & -325354.810 & 37874.047 & (-332906.876, -317802.744) & Non-VLSN \\
\hline
\multirow{4}{*}{\textbf{VLSNMS}} & \textbf{Greedy} & 4  & 30 & 457678.140 & 293766.388 & (399101.268, 516255.012) & VLSN \\
 & \textbf{GA1} & 4  & 30 & 7632962.370 & 322434.530 & (7568669.085, 7697255.655) & VLSN \\
 & \textbf{GA2} & 4  & 30 & 7592835.890 & 365974.022 & (7519860.852, 7665810.928) & VLSN \\
 & \textbf{Gurobi} & 4  & 30 & -169555.910 & 10206.322 & (-171591.046, -167520.774) & Non-VLSN \\
\hline
\multirow{4}{2mm}{\textbf{Grid-VLSN}} & \textbf{Greedy} & 4  & 30 & 465655.330 & 295055.791 & (406821.352, 524489.308) & VLSN \\
 & \textbf{GA1} & 4  & 30 & 7640939.560 & 323539.582 & (7576425.928, 7705453.192) & VLSN \\
 & \textbf{GA2} & 4  & 30 & 7600813.080 & 364738.028 & (7528084.499, 7673541.661) & VLSN \\
 & \textbf{Gurobi} & 4  & 30 & -161578.720 & 10410.772 & (-163654.623, -159502.817) & Non-VLSN \\
\hline
\multirow{4}{2mm}{\textbf{Greedy-VLSN}} & \textbf{Greedy} & 4  & 30 & 503226.990 & 291500.030 & (85368.942, 509448384.965) & VLSN \\
 & \textbf{GA1} & 4  & 30 & 7678511.220 & 324933.084 & (7613719.725, 7743302.715) & VLSN \\
 & \textbf{GA2} & 4  & 30 & 7638384.740 & 364695.841 & (7565664.571, 7711104.909) & VLSN \\
 & \textbf{Gurobi} & 4  & 30 & -124007.060 & 35641.249 & (-131113.907, -116900.213) & Non-VLSN \\
\hline
\hline
%\end{tabular}
%\end{table}
\end{longtable}
\end{center}

Column \textbf{$A_1$: VLSN} contains a name of VLSN version. Column \textbf{$A_2$: Non-VLSN} lists a name of non-VLSN algorithm that is compared with the VLSN search in the same row. Columns $A_{1,2}$, $D_{1,2}$, and \textbf{95\% C.I.} display the sample mean, the sample standard deviation, and the 95\% %2-$\sigma$ 
confidence interval for $y_1 - y_2$.

}%%%%%%%%%%%%%%%%%%%%%%% END COLOR BLUE

{\color{blue}
\subsection{Illustration of application to multipartite entity resolution problem}~\label{sec:real-exp}
Current experiment presents illustration of application of proposed algorithms to multipartite entity resolution problem. Experiment settings and results are displayed in table \ref{exp-2}. We have collected list of 2610 highly cited researcher names from (http://www.webometrics.info/en/node/58) and selected subset of size \textbf{N} of their names. Next, we created three identical datasets $D_1$, $D_2$, and $D_3$. Column \textbf{e} denotes \textit{error}: probability under which random letter of any item in $D_1$ or $D_2$ would be changed to random letter. Part of distorted dataset is shown in table \ref{src-d}. We have generated input matrix as per formulation \ref{eq:delta}, using trigram similarity as similarity function, and performed greedy search as in  algorithm \ref{alg:greedy}, and then improved it using VLSN with greedy start (alg. \ref{alg:VLSN.OP}).
Algorithms return solutions as set of 3-tuples $( (d_1^1, d_2^1, d_3^1), ... , (d_1^N, d_2^N, d_3^N) )$, meaning that $d_1^1$, $d_2^1$, and $d_3^1$ denote the same entity. We define 3-tuple as correctly matched, if all its elements correspond to the record for the same entity, otherwise the 3-tuple is matched incorrectly.
We define \textit{recall} of the algorithm as the ratio of correctly matched 3-tuples to the total number of 3-tuples. Columns \textbf{R(G)} and \textbf{R(V\_G)} denote recall for greedy and for VLSN-greedy algorithms respectively. Column \textbf{$|V|$} denotes number of nodes in the resulting graph.
Results of this experiment show that recall for VLSN search improves over greedy search.

\begin{table}[]
\centering
\caption{part of source data for e = 40}
\label{src-d}
\begin{tabular}{|l|l|l|}
\hline
$D_1$             & $D_2$             & $D_3$             \\ \hline
Sigmund Freud    & Sigmund Freud    & SigmundxFreuv    \\ \hline
Graham Colditz   & srShamNConditz   & GrahaGOColditu   \\ \hline
Ronald C Kessler & RLnald C KmsslXr & pocalt C KEssler \\ \hline
JoAnn E Manson   & hoAnn AhYansan   & JoAnniE Manson   \\ \hline
Shizuo Akira     & ShMzuI Tkila     & zhizuoyAkCra     \\ \hline
\end{tabular}
\end{table}

\begin{table}[]
\centering
\caption{Results of application of MAP to multipartite entity resolution problem}
\label{exp-2}
\begin{tabular}{|l|l|l|l|l|l|l|l|}
\hline
\textbf{\#} & \textbf{N} & \textbf{e} & \textbf{G} & \textbf{V\_G} & \textbf{$|V|$} & \textbf{R(G)} & \textbf{R(V\_G)} \\ \hline
1           & 10         & 0          & 10         & 10            & 1          & 1.0           & 1.0              \\ \hline
2           & 10         & 50         & 6          & 10            & 4          & 0.6           & 1.0              \\ \hline
3           & 20         & 20         & 20         & 20            & 1          & 1.0           & 1.0              \\ \hline
4           & 20         & 30         & 17         & 20            & 2          & 0.85          & 1.0              \\ \hline
5           & 40         & 20         & 40         & 40            & 1          & 1.0           & 1.0              \\ \hline
6           & 20         & 50         & 16         & 20            & 4          & 0.8           & 1.0              \\ \hline
7           & 50         & 10         & 50         & 50            & 1          & 1.0           & 1.0              \\ \hline
8           & 10         & 20         & 10         & 10            & 1          & 1.0           & 1.0              \\ \hline
9           & 20         & 10         & 20         & 20            & 1          & 1.0           & 1.0              \\ \hline
10          & 20         & 40         & 20         & 20            & 1          & 1.0           & 1.0              \\ \hline
11          & 10         & 40         & 6          & 8             & 4          & 0.6           & 0.8              \\ \hline
12          & 20         & 0          & 20         & 20            & 1          & 1.0           & 1.0              \\ \hline
13          & 10         & 30         & 8          & 10            & 2          & 0.8           & 1.0              \\ \hline
14          & 10         & 10         & 10         & 10            & 1          & 1.0           & 1.0              \\ \hline
15          & 40         & 10         & 40         & 40            & 1          & 1.0           & 1.0              \\ \hline
16          & 50         & 20         & 48         & 48            & 2          & 0.96          & 0.96             \\ \hline
17          & 40         & 30         & 31         & 38            & 4          & 0.775         & 0.95             \\ \hline
18          & 40         & 40         & 30         & 37            & 7          & 0.75          & 0.925            \\ \hline
19          & 100        & 10         & 100        & 100           & 1          & 1.0           & 1.0              \\ \hline
20          & 50         & 30         & 36         & 50            & 5          & 0.72          & 1.0              \\ \hline
21          & 100        & 20         & 96         & 98            & 2          & 0.96          & 0.98             \\ \hline
22          & 50         & 40         & 38         & 46            & 13         & 0.76          & 0.92             \\ \hline
23          & 40         & 50         & 22         & 26            & 25         & 0.55          & 0.65             \\ \hline
24          & 50         & 50         & 20         & 28            & 13         & 0.4           & 0.56             \\ \hline
25          & 100        & 30         & 72         & 96            & 13         & 0.72          & 0.96             \\ \hline
26          & 300        & 10         & 289        & 300           & 4          & 0.96   & 1.0              \\ \hline
27          & 100        & 40         & 54         & 71            & 35         & 0.54          & 0.71             \\ \hline
28          & 100        & 50         & 31         & 47            & 83         & 0.31          & 0.47             \\ \hline
29          & 300        & 20         & 269        & 300           & 9          & 0.89   & 1.0              \\ \hline
30          & 300        & 30         & 207        & 276           & 20         & 0.69          & 0.92             \\ \hline
31          & 300        & 40         & 131        & 201           & 82         & 0.43   & 0.67             \\ \hline
\end{tabular}
\end{table}
}

\section{Conclusions and further research}\label{sec:end}
Current paper describes mathematical derivation of multidimensional assignment problem and its application to multipartite entity matching. Exact solution of MAP is NP-hard; in current paper we present two heuristic approaches to solve the problem. Because the new application to multipartite entity resolution requires solving problems with large cardinality, exact approaches become computationally infeasible. Heuristic and metaheuristic approaches are more scalable; but they require intelligent strategies for exploration of vast solution set. Therefore, understanding which multistart approaches should be used in exploration phase of a search become important. We add to this under-investigated area by studying three alternative ways to restart VLSN search algorithm, including random restart, grid-based restart and greedy restart.
%Once applied to entity resolution, MAP solution algorithms demonstrate increased recall.

Limitation of current approach is necessity to have entire matrix with comparison filled with values; it means, that all elements of all datasets should be compared to each other. Due to practical limitations that cannot be performed for large datasets. We aim at resolving this limitation by reformulating MAP taking into account blocking, or Top-K similarity queries. Next, in current experiments matrices are generated using uniform random distribution. We believe that in real data, when matrices are constructed using comparisons, distributions will be other than uniform random, and it would lead to different properties of solutions.

\section*{Acknowledgment}
A. Kammerdiner was supported by the AFRL (National Research Council Fellowship). The work of A. Semenov was funded in part by the AFRL European Office of Aerospace Research and Development (Grant FA9550-17-1-0030, University of Jyvaskyla, Finland).

\section*{Data availability statement}

All data generated or analysed during this study are included in this published article.

\bibliographystyle{abbrv}
\bibliography{main} 

\begin{thebibliography}{10}

\bibitem{icwsm_2009}
E.~Adar, M.~Hurst, T.~Finin, N.~S. Glance, N.~Nicolov, and B.~L. Tseng,
  editors.
\newblock {\em Proceedings of the Third International Conference on Weblogs and
  Social Media, {ICWSM} 2009, San Jose, California, USA, May 17-20, 2009}. The
  {AAAI} Press, 2009.

\bibitem{arbib1999three}
C.~Arbib, D.~Pacciarelli, and S.~Smriglio.
\newblock A three-dimensional matching model for perishable production
  scheduling.
\newblock {\em Discrete Applied Mathematics}, 92(1):1--15, 1999.

\bibitem{balas1983traffic}
E.~Balas and P.~R. Landweer.
\newblock Traffic assignment in communication satellites.
\newblock {\em Operations Research Letters}, 2(4):141--147, 1983.

\bibitem{benjelloun_swoosh_2009}
O.~Benjelloun, H.~Garcia-Molina, D.~Menestrina, Q.~Su, S.~E. Whang, and
  J.~Widom.
\newblock Swoosh: {A} {Generic} {Approach} to {Entity} {Resolution}.
\newblock {\em The VLDB Journal}, 18(1):255--276, Jan. 2009.

\bibitem{brizan_survey_2006}
D.~Brizan and A.~Tansel.
\newblock A {Survey} of {Entity} {Resolution} and {Record} {Linkage}
  {Methodologies}.
\newblock {\em Communications of the IIMA}, 6(3):41--50, 2006.

\bibitem{burkard2012assignment}
R.~Burkard, M.~Dell'Amico, and S.~Martello.
\newblock {\em Assignment problems, revised reprint}, volume 106.
\newblock Siam, 2012.

\bibitem{burkard1999linear}
R.~E. Burkard and E.~Cela.
\newblock Linear assignment problems and extensions.
\newblock In {\em Handbook of combinatorial optimization}, pages 75--149.
  Springer, 1999.

\bibitem{Chu_VLDB_2016}
X.~Chu, I.~F. Ilyas, and P.~Koutris.
\newblock Distributed data deduplication.
\newblock {\em Proc. VLDB Endow.}, 9(11):864--875, July 2016.

\bibitem{crama1997assembly}
Y.~Crama, O.~E. Flippo, J.~Van~de Klundert, and F.~C. Spieksma.
\newblock The assembly of printed circuit boards: A case with multiple machines
  and multiple board types.
\newblock {\em European journal of operational research}, 98(3):457--472, 1997.

\bibitem{crama2012production}
Y.~Crama, A.~G. Oerlemans, and F.~C. Spieksma.
\newblock {\em Production planning in automated manufacturing}.
\newblock Springer Science \& Business Media, 2012.

\bibitem{elmagarmid_duplicate_2007}
A.~K. Elmagarmid, P.~G. Ipeirotis, and V.~S. Verykios.
\newblock Duplicate {Record} {Detection}: {A} {Survey}.
\newblock {\em IEEE Transactions on Knowledge and Data Engineering},
  19(1):1--16, Jan. 2007.

\bibitem{firmani_online_2016}
D.~Firmani, B.~Saha, and D.~Srivastava.
\newblock Online {Entity} {Resolution} {Using} an {Oracle}.
\newblock {\em Proc. VLDB Endow.}, 9(5):384--395, Jan. 2016.

\bibitem{frieze1981algorithm}
A.~Frieze and J.~Yadegar.
\newblock An algorithm for solving 3-dimensional assignment problems with
  application to scheduling a teaching practice.
\newblock {\em Journal of the operational research society}, 32(11):989--995,
  1981.

\bibitem{garey1979computers}
M.~R. Garey and D.~S. Johnson.
\newblock {\em Computers and intractability}, volume 174.
\newblock freeman San Francisco, 1979.

\bibitem{gilbert1987algorithm}
K.~C. Gilbert and R.~B. Hofstra.
\newblock An algorithm for a class of three-dimensional assignment problems
  arising in scheduling applications.
\newblock {\em IIE transactions}, 19(1):29--33, 1987.

\bibitem{gilbert1988multidimensional}
K.~C. Gilbert and R.~B. Hofstra.
\newblock Multidimensional assignment problems.
\newblock {\em Decision Sciences}, 19(2):306--321, 1988.

\bibitem{Gokhale_sigmod_2014}
C.~Gokhale, S.~Das, A.~Doan, J.~F. Naughton, N.~Rampalli, J.~Shavlik, and
  X.~Zhu.
\newblock Corleone: Hands-off crowdsourcing for entity matching.
\newblock In {\em Proceedings of the 2014 ACM SIGMOD International Conference
  on Management of Data}, SIGMOD '14, pages 601--612, New York, NY, USA, 2014.
  ACM.

\bibitem{guo_2010_vldb}
S.~Guo, X.~L. Dong, D.~Srivastava, and R.~Zajac.
\newblock Record linkage with uniqueness constraints and erroneous values.
\newblock {\em Proc. VLDB Endow.}, 3(1-2):417--428, Sept. 2010.

\bibitem{gutin2008worst}
G.~Gutin, B.~Goldengorin, and J.~Huang.
\newblock Worst case analysis of max-regret, greedy and other heuristics for
  multidimensional assignment and traveling salesman problems.
\newblock {\em Journal of Heuristics}, 14(2):169--181, 2008.

\bibitem{he_relationship_2016}
J.~He, H.~Liu, R.~Y.~K. Lau, and J.~He.
\newblock Relationship {Identification} {Across} {Heterogeneous} {Online}
  {Social} {Networks}.
\newblock {\em Computational Intelligence}, Jan. 2016.

\bibitem{Helbing2011}
D.~Helbing and S.~Balietti.
\newblock From social data mining to forecasting socio-economic crises.
\newblock {\em The European Physical Journal Special Topics}, 195(1):3, 2011.

\bibitem{hilton1980reconstruction}
A.~Hilton.
\newblock The reconstruction of latin squares with applications to school
  timetabling and to experimental design.
\newblock In {\em Combinatorial Optimization II}, pages 68--77. Springer, 1980.

\bibitem{Jain_2013}
P.~Jain, P.~Kumaraguru, and A.~Joshi.
\newblock @i seek 'fb.me': Identifying users across multiple online social
  networks.
\newblock In {\em Proceedings of the 22Nd International Conference on World
  Wide Web}, WWW '13 Companion, pages 1259--1268, New York, NY, USA, 2013. ACM.

\bibitem{kammerdiner2015ranking}
A.~Kammerdiner.
\newblock Ranking risk exposures for situational surveillance of falls with
  sensors.
\newblock {\em Operations Research for Health Care}, 7:132--137, 2015.

\bibitem{kammerdiner2007characteristics}
A.~Kammerdiner, P.~Krokhmal, and P.~Pardalos.
\newblock Characteristics of the distribution of hamming distance values
  between multidimensional assignment problem solutions.
\newblock {\em Advances in Cooperative Control and Optimization}, pages
  339--352, 2007.

\bibitem{kammerdiner2017very}
A.~Kammerdiner and C.~F. Vaughan.
\newblock Very large-scale neighborhood search for the multidimensional
  assignment problem.
\newblock In S.~Butenko, P.~M. Pardalos, and V.~Shylo, editors, {\em
  Optimization Methods and Applications}. Springer, 2017.

\bibitem{kammerdiner2008multidimensional}
A.~R. Kammerdiner.
\newblock Multidimensional assignment problem multidimensional assignment
  problem.
\newblock In {\em Encyclopedia of Optimization}, pages 2396--2402. Springer,
  2008.

\bibitem{kammerdiner2019data}
A.~R. Kammerdiner and A.~N. Guererro.
\newblock Data-driven combinatorial optimization for sensor-based assessment of
  near falls.
\newblock {\em Annals of Operations Research}, 276(1-2):137--153, 2019.

\bibitem{kammerdiner2009application}
A.~R. Kammerdiner, A.~Mucherino, and P.~M. Pardalos.
\newblock Application of monkey search meta-heuristic to solving instances of
  the multidimensional assignment problem.
\newblock In {\em Optimization and Cooperative Control Strategies}, pages
  385--397. Springer, 2009.

\bibitem{karapetyan2011local}
D.~Karapetyan and G.~Gutin.
\newblock Local search heuristics for the multidimensional assignment problem.
\newblock {\em Journal of Heuristics}, 17(3):201--249, 2011.

\bibitem{karp1972reducibility}
R.~M. Karp.
\newblock Reducibility among combinatorial problems.
\newblock In {\em Complexity of computer computations}, pages 85--103.
  Springer, 1972.

\bibitem{Kopcke_VLDB_2010}
H.~K\"{o}pcke, A.~Thor, and E.~Rahm.
\newblock Evaluation of entity resolution approaches on real-world match
  problems.
\newblock {\em Proc. VLDB Endow.}, 3(1-2):484--493, Sept. 2010.

\bibitem{krokhmal2011optimality}
P.~A. Krokhmal.
\newblock On optimality of a polynomial algorithm for random linear
  multidimensional assignment problem.
\newblock {\em Optimization Letters}, 5(1):153--164, 2011.

\bibitem{Li_sigmod_2015}
F.~Li, M.~L. Lee, W.~Hsu, and W.-C. Tan.
\newblock Linking temporal records for profiling entities.
\newblock In {\em Proceedings of the 2015 ACM SIGMOD International Conference
  on Management of Data}, SIGMOD '15, pages 593--605, New York, NY, USA, 2015.
  ACM.

\bibitem{nguyen2014solving}
D.~M. Nguyen, H.~A. Le~Thi, and T.~P. Dinh.
\newblock Solving the multidimensional assignment problem by a cross-entropy
  method.
\newblock {\em Journal of Combinatorial Optimization}, 27(4):808--823, 2014.

\bibitem{Papadakis_VLDB_2016}
G.~Papadakis, J.~Svirsky, A.~Gal, and T.~Palpanas.
\newblock Comparative analysis of approximate blocking techniques for entity
  resolution.
\newblock {\em Proc. VLDB Endow.}, 9(9):684--695, May 2016.

\bibitem{pasiliao2010local}
E.~L. Pasiliao~Jr.
\newblock Local neighborhoods for the multidimensional assignment problem.
\newblock In {\em Dynamics of information systems}, pages 353--371. Springer,
  2010.

\bibitem{pierskalla1967tri}
W.~P. Pierskalla.
\newblock The tri-substitution method for the three-dimensional assignment
  problem.
\newblock {\em CORS Journal}, 5:71--81, 1967.

\bibitem{pierskalla1968letter}
W.~P. Pierskalla.
\newblock Letter to the editor—the multidimensional assignment problem.
\newblock {\em Operations Research}, 16(2):422--431, 1968.

\bibitem{poore1993data}
A.~Poore, N.~Rijavec, M.~Liggins, and V.~Vannicola.
\newblock Data association problems posed as multidimensional assignment
  problems: problem formulation.
\newblock In {\em Optical Engineering and Photonics in Aerospace Sensing},
  pages 552--563. International Society for Optics and Photonics, 1993.

\bibitem{puglisi_web_2016}
S.~Puglisi, D.~Rebollo-Monedero, and J.~Forné.
\newblock On {Web} user tracking: {How} third-party http requests track users'
  browsing patterns for personalised advertising.
\newblock In {\em 2016 {Mediterranean} {Ad} {Hoc} {Networking} {Workshop}
  ({Med}-{Hoc}-{Net})}, pages 1--6, June 2016.

\bibitem{pusztaszeri2000nonlinear}
J.-F. Pusztaszeri.
\newblock The nonlinear assignment problem in experimental high energy physics.
\newblock In {\em Nonlinear Assignment Problems}, pages 55--89. Springer, 2000.

\bibitem{pusztaszeri1996tracking}
J.-F. Pusztaszeri, P.~E. Rensing, and T.~M. Liebling.
\newblock Tracking elementary particles near their primary vertex: a
  combinatorial approach.
\newblock {\em Journal of Global Optimization}, 9(1):41--64, 1996.

\bibitem{riederer_linking_2016}
C.~Riederer, Y.~Kim, A.~Chaintreau, N.~Korula, and S.~Lattanzi.
\newblock Linking {Users} {Across} {Domains} with {Location} {Data}: {Theory}
  and {Validation}.
\newblock In {\em Proceedings of the 25th {International} {Conference} on
  {World} {Wide} {Web}}, {WWW} '16, pages 707--719, Republic and Canton of
  Geneva, Switzerland, 2016. International World Wide Web Conferences Steering
  Committee.

\bibitem{sagi_sigmod_2016}
T.~Sagi, A.~Gal, O.~Barkol, R.~Bergman, and A.~Avram.
\newblock Multi-source uncertain entity resolution at yad vashem: Transforming
  holocaust victim reports into people.
\newblock In {\em Proceedings of the 2016 International Conference on
  Management of Data}, SIGMOD '16, pages 807--819, New York, NY, USA, 2016.
  ACM.

\bibitem{ijwet2013}
A.~Semenov and J.~Veijalainen.
\newblock A modelling framework for social media monitoring.
\newblock {\em International Journal of Web Engineering and Technology},
  8(3):217--249, 2013.
\newblock PMID: 57226.

\bibitem{kai_shu_2016}
K.~Shu, S.~Wang, J.~Tang, R.~Zafarani, and H.~Liu.
\newblock User identity linkage across online social netwroks: A review.
\newblock In {\em To appear in SIGKDD Explorations}, To appear in SIGKDD
  Explorations, 2016.

\bibitem{10.1007/s10898-013-0105-7}
R.~M. Silva, M.~G. Resende, and P.~M. Pardalos.
\newblock Finding multiple roots of a box-constrained system of nonlinear
  equations with a biased random-key genetic algorithm.
\newblock {\em J. of Global Optimization}, 60(2):289–306, Oct. 2014.

\bibitem{vogiatzis2014graph}
C.~Vogiatzis, E.~L. Pasiliao, and P.~M. Pardalos.
\newblock Graph partitions for the multidimensional assignment problem.
\newblock {\em Computational Optimization and Applications}, 58(1):205--224,
  2014.

\bibitem{wang_crowder_2012}
J.~Wang, T.~Kraska, M.~J. Franklin, and J.~Feng.
\newblock {CrowdER}: {Crowdsourcing} {Entity} {Resolution}.
\newblock {\em Proc. VLDB Endow.}, 5(11):1483--1494, July 2012.

\bibitem{Ye_CIKM_2015}
T.~Ye and H.~W. Lauw.
\newblock Structural constraints for multipartite entity resolution with markov
  logic network.
\newblock In {\em Proceedings of the 24th ACM International on Conference on
  Information and Knowledge Management}, CIKM '15, pages 1691--1694, New York,
  NY, USA, 2015. ACM.

\bibitem{zhang_principled_2015}
D.~Zhang, B.~I.~P. Rubinstein, and J.~Gemmell.
\newblock Principled {Graph} {Matching} {Algorithms} for {Integrating}
  {Multiple} {Data} {Sources}.
\newblock {\em IEEE Transactions on Knowledge and Data Engineering},
  27(10):2784--2796, Oct. 2015.

\bibitem{zhang_multiple_2015}
J.~Zhang and P.~S. Yu.
\newblock Multiple {Anonymized} {Social} {Networks} {Alignment}.
\newblock In {\em 2015 {IEEE} {International} {Conference} on {Data} {Mining}},
  pages 599--608, Nov. 2015.

\bibitem{Zhou_2016}
X.~Zhou, X.~Liang, H.~Zhang, and Y.~Ma.
\newblock Cross-platform identification of anonymous identical users in
  multiple social media networks.
\newblock {\em IEEE Trans. on Knowl. and Data Eng.}, 28(2):411--424, Feb. 2016.

\bibitem{zhou1997confidence}
X.-H. ZHOU and S.~Gao.
\newblock Confidence intervals for the log-normal mean.
\newblock {\em Statistics in medicine}, 16(7):783--790, 1997.

\end{thebibliography}
\end{document}